\documentclass[10pt]{iopart}
\usepackage{graphicx}
\usepackage{cite}
\usepackage{eqnarray}
\usepackage[colorlinks,
linkcolor=blue,
anchorcolor=blue,
citecolor=blue,
urlcolor=blue
]{hyperref}

\begin{document}
\title[]{Non-linear MHD modelling of shattered pellet injection in ASDEX Upgrade}

\author{W. Tang$^1$, M. Hoelzl$^1$, M. Lehnen$^{2}$\footnote[1]{Deceased}, D. Hu$^3$, F. J. Artola$^2$, P. Halldestam$^1$, P. Heinrich$^1$, S. Jachmich$^2$, E. Nardon$^4$,\\ G. Papp$^1$, A. Patel$^1$, the ASDEX Upgrade Team$^a$, the EUROfusion Tokamak Exploitation Team$^b$ and the JOREK Team$^c$}         

\address{$^1$Max Planck Institute for Plasma Physics, Garching, Germany\\
         $^2$ITER Organization, Saint-Paul-Lez-Durance, France\\
         $^3$Beihang University, Beijing, China\\
         $^4$CEA/IRFM, Saint-Paul-Lez-Durance, France\\
         $^a$See the author list of H. Zohm et al. 2024 \textit{Nucl. Fusion} \textbf{64} 112001\\
         $^b$See the author list of E. Joffrin et al. 2024 \textit{Nucl. Fusion} \textbf{64} 112019\\
         $^c$See the author list of M. Hoelzl et al. 2024 \textit{Nucl. Fusion} \textbf{64} 112016
         }
\ead{weikang.tang@ipp.mpg.de}
\vspace{10pt}

\begin{abstract}
    Shattered pellet injection (SPI) is selected for the disruption mitigation system in ITER, due to deeper penetration, expected assimilation efficiency and prompt material delivery. This article describes non-linear magnetohydrodynamic (MHD) simulations of SPI in the ASDEX Upgrade tokamak to test the mitigation efficiency of different injection parameters for neon-doped deuterium pellets using the JOREK code. The simulations are executed as fluid simulations, while additional marker particles are used to evolve the charge state distribution and radiation property of impurities based on OpenADAS atomic data, i.e., a collisional-radiative model is used. Neon fraction scans between 0 - 10\% are performed. Numerical results show that the thermal quench (TQ) occurs in two stages. In the first stage, approximately half of the thermal energy is abruptly lost, primarily through convective and conductive transport in the stochastic fields. This stage is relatively independent of the neon fraction. In the second stage, where the majority of the remaining thermal energy is lost, radiation plays a dominant role. In case of pure deuterium injection, this second stage may not occur at all. A larger fraction ($\sim $20\%) of the total material in the pellet is assimilated in the plasma for low neon fraction pellets ($\leq 0.12\%$) due to the full thermal collapse of the plasma occurring later than in high neon fraction scenarios. Nevertheless, the total number of assimilated neon atoms increases with increasing neon fraction. The effects of fragment size and penetration speed are then numerically studied, showing that slower and smaller fragments promote edge cooling and the formation of a cold front. Faster fragments result in shorter TQ duration and higher assimilation as they reach the hotter plasma regions quicker.
\end{abstract}

\vspace{2pc}



\ioptwocol

\section{Introduction}\label{sec1}
    Unmitigated major disruptions~\cite{Boozer2012} could potentially damage the plasma facing components and the structural integrity of large fusion devices due to heat loads, electromagnetic forces, and runaway electrons (REs)~\cite{Breizman_2019}. To effectively mitigate all these above consequences, the ITER disruption mitigation system (DMS) is designed based on the shattered pellet injection (SPI)~\cite{DMSIAEA}. SPI is a technique that can actively shut down the `unhealthy' plasma, by injecting a cryogenic pellet composed of pure protium or deuterium, or a mixture also containing neon~\cite{SPIORNL,Gebhart_2021}. The injected protium/deuterium aids in suppressing REs, while the ablated impurities can radiate away the stored energy uniformly, hence reducing the localized heat loads on the plasma facing components. To develop a thorough understanding of SPI, essential for extrapolating its performance to future machines like ITER, SPI systems are installed on tokamak devices worldwide, including DIII-D~\cite{Commaux_2016}, JET~\cite{Jachmich_2022}, KSTAR~\cite{KSTARSPI}, J-TEXT~\cite{JTEXTSPI}, HL-2A~\cite{HLSPI}, EAST~\cite{EASTSPI}, and ASDEX Upgrade (AUG)~\cite{Dibon}. On the modelling side, significant efforts are being undertaken to study the material assimilation, RE mitigation and magnetohydrodynamic (MHD) response after injection. Various codes, including the 3D MHD codes NIMROD~\cite{NIMRODSPI}, M3D-C1~\cite{Lyons_2019} and JOREK~\cite{Lee_2024}, as well as lower dimensional codes like INDEX~\cite{Matsuyama_2022,ansh} and DREAM~\cite{Vallhagen_2022}, are being employed in these studies. 

    To determine the optimum fragment size, speed, and neon mixture fraction for the ITER DMS, the AUG tokamak has installed a flexible SPI system with three independent injection lines, allowing for multiple simultaneous injections with options of different shatter heads~\cite{HEINRICH2024114576}. After extensive lab tests and commissioning~\cite{Dibon,HEINRICH2024114576}, SPI experiments with different shattering scenarios were carried out in AUG during the 2022 campaign~\cite{paul2024arxiv}. Due to precise lab characterization of the fragment cloud~\cite{tobias,johannes2024} and an excellent bolometry coverage~\cite{paul2024arxiv}, experiments in AUG and simulations for AUG scenarios can substantially advance the understanding of the involved processes and can help to robustly validate the models in view of reliable predictive simulations for ITER. 
    
    The 3D non-linear MHD studies addressed here aim to gain a deeper understanding of complex physical phenomena by fully accounting for various self-consistent processes that are simplified or neglected in lower-dimensional models. These studies, although computationally more expensive, are crucial because they allow for the detailed inclusion of more physical processes, such as relaxation of the temperature profile by stochastic transport, current profile evolution due to resistive decay in the presence of the very large temperature variations~\cite{Nardon_2023}, and plasmoid drift dynamics originating in the vertical motion of the electrons and ions in the inhomogeneous magnetic field~\cite{Pegourie_2007,plasmoid_kong}. 

    The detailed nature of 3D non-linear studies also allows for direct comparisons with experimental data via synthetic diagnostics, e.g., regarding localized density measurements~\cite{Kong_2024,Matthias2020}, wall heat deposition~\cite{Hu_2024}, and electromagnetic forces in a 3D context~\cite{Schwarz_2023}, which are essential for understanding real-world scenarios. However, despite the advancements these studies provide, there is still much work to be done in interpreting experimental results and validating these models against empirical data. This is where systematic parameter scans become invaluable, especially given the challenges associated with full thermal quench (TQ) simulations.
    
    In the present article, 3D non-linear SPI simulations for an AUG H-mode scenario are carried out using the JOREK code~\cite{Huysmans2007,jorek2024}. Our simulations are focusing on the full thermal quench as well as the early current quench (CQ) phase. Mitigation efficiency of different injection parameters for neon-doped deuterium pellets is numerically studied. Detailed analysis and variations of the injection parameters reveal fundamental dependencies. The rest of the paper will be structured as follows: In section~\ref{sec2}, the governing reduced MHD equations are introduced, as well as the fragment configuration describing the shattering process. In section~\ref{sec3}, the effectiveness of SPI under various injection parameters is investigated. Comparisons of general trend between simulations and experiments are conducted using synthetic diagnostics.

\section{Simulation model}\label{sec2}
    \subsection{The governing equations}\label{equation}
        For our studies we employ the visco-resistive two-temperature reduced MHD model of JOREK in the realistic tokamak geometry as briefly summarized in the following.
        \begin{eqnarray}
        \frac{\partial \psi}{\partial t}=\eta \Delta^* \psi-R\{u, \psi\}-F_0 \frac{\partial u}{\partial \phi},
        \end{eqnarray} 
    
        \begin{eqnarray}
        R \nabla &\cdot\left[R^2\left(\rho \nabla_{p o l} \frac{\partial u}{\partial t}+\nabla_{p o l} u \frac{\partial \rho}{\partial t}\right)\right]= \nonumber\\
                 & \frac{1}{2}\left\{R^2\left|\nabla_{p o l} u\right|^2, R^2 \rho\right\}+\left\{R^4 \rho \omega, u\right\} \nonumber\\
                 & -R \nabla \cdot\left[R^2 \nabla_{p o l} u \nabla \cdot(\rho \mathbf{v})\right]+\{\psi, j\}-\frac{F_0}{R} \frac{\partial j}{\partial \phi} \nonumber\\
                 & +\left\{P, R^2\right\}+R \mu \nabla_{p o l}^2 \omega, 
        \end{eqnarray}
    
        \begin{eqnarray}
        B^2 \frac{\partial}{\partial t}\left(\rho v_{\|}\right)=&-\frac{1}{2} \rho \frac{F_0}{R^2} \frac{\partial}{\partial \phi}\left(v_{\|} B\right)^2-\frac{\rho}{2 R}\left\{B^2 v_{\|}^2, \psi\right\} \nonumber\\
                                                                &-\frac{F_0}{R^2} \frac{\partial P}{\partial \phi}+\frac{1}{R}\{\psi, P\} \nonumber\\
                                                                &-B^2 \nabla \cdot(\rho \mathbf{v}) v_{\|}+B^2 \mu_{\|} \nabla_{p o l}^2 v_{\|},
        \end{eqnarray}

        \begin{eqnarray}
        \frac{\partial}{\partial t} \rho=&-\nabla \cdot(\rho \mathbf{v})+\nabla \cdot\left[D_D \nabla\left(\rho-\rho_{i m p}\right)\right] \nonumber\\
                                         &+\nabla \cdot\left(D_{i m p} \nabla \rho_{i m p}\right)+S_D+S_{i m p},    
        \end{eqnarray}

        \begin{eqnarray}
        \frac{\partial}{\partial t} \rho_{i m p}=-\nabla \cdot\left(\rho_{i m p} \mathbf{v}\right) +\nabla \cdot\left(D_{i m p} \nabla \rho_{i m p}\right)+S_{i m p},
        \end{eqnarray}

        \begin{eqnarray}
        \frac{\partial}{\partial t} P_i= &-\mathbf{v} \cdot \nabla P_i-\gamma P_i \nabla \cdot \mathbf{v} \nonumber\\
                                         &+\nabla \cdot\left(\kappa_{\perp} \nabla_{\perp} T_i+\kappa_{i, \|} \nabla_{\|} T_i\right) \nonumber\\
                                         &+\frac{\gamma-1}{2} \mathbf{v} \cdot \mathbf{v}\left(S_{D}+S_{i m p}\right)+(\gamma-1) \mu_{\|} \nonumber\\
                                         &\times\left[\nabla_{p o l}\left(v_{\|} B\right)\right]^2+\left(n_{D}+n_{i m p}\right)\left(\partial_t T_i\right)_{c, e},
        \end{eqnarray}

        \begin{eqnarray}
        \frac{\partial}{\partial t} P_e= &-\mathbf{v} \cdot \nabla P_e-\gamma P_e \nabla \cdot \mathbf{v} \nonumber\\
                                         &+\nabla \cdot\left(\kappa_{\perp} \nabla_{\perp} T_e+\kappa_{e, \|} \nabla_{\|} T_e\right) \nonumber\\
                                         &+\frac{\gamma-1}{R^2} \eta j^2 -(\gamma-1)\left(P_{rad}+P_{ion}\right) \nonumber\\
                                         &+n_e\left(\partial_t T_e\right)_{c, i} .
        \end{eqnarray}

        The main scalar fields of interest in the reduced MHD system are, the poloidal magnetic flux $\psi$, the current density $j$, the vorticity $\omega$, the stream function~$u$, the parallel velocity $v_{\|}$, the total mass density $\rho$, the impurity mass density $\rho_{imp}$, the ion/electron pressure $P_{i/e}$ and the ion/electron temperature $T_{i/e}$. The diffusive coefficients $\eta$, $\mu$, $\mu_{\|}$, $D_D$, $D_{imp}$, $\kappa_{\perp}$, $\kappa_{i,\|}$, $\kappa_{e,\|}$ are respectively the resistivity, perpendicular viscosity, parallel viscosity, particle diffusivity for main ions, particle diffusivity for impurity, perpendicular thermal conductivity, parallel thermal conductivity for ions and electrons. The source terms $S_D$ and $S_{imp}$ are the SPI ablation sources for main ion and impurity species, respectively.   
        The Poisson bracket is defined as $\{f, g\}=R(\nabla f\times\nabla g)\cdot\nabla\phi$. Detailed simulation parameters are given in section~\ref{setup}.

        A coronal equilibrium distribution of impurity charge states had been utilized in previous JOREK SPI simulations~\cite{Hu_2021}. To obtain a more accurate evolution of charge states and radiation, the impurity charge states are self-consistently evolved in time using OpenADAS atomic coefficients~\cite{Hu2021b}. Here, marker particles are employed to carry the charge state evolution, while all other quantities are treated in the fluid picture. Markers are created according to the impurity ablation source $S_{imp}$ and are convected with the fluid velocity field. The ionization and recombination processes are calculated according to the local temperature and density fields. The atomic processes taking into account the charge state distribution carried by the makers provides the necessary information to calculate the radiation power density $P_{rad}$ and ionization power density $P_{ion}$. 

\begin{table*}[htb!]
\centering
\caption{The SPI injection parameters for the different simulations performed in this study. Naming convention: \textbf{L}arge/\textbf{M}edium/\textbf{S}mall \textbf{F}ragment as well as \textbf{F}ull/\textbf{H}alf \textbf{V}elocity and the neon atomic mixture ratio in the pellet. * for the half speed case (222 m/s), to isolate the effect of the penetration speed, we maintain the same fragment sizes as in LF\_FV\_Ne10 and MF\_FV\_Ne10, while manually reducing the velocities of the fragments to half of their original values. It is equivalent to keeping the $v_{\perp}$ while reducing the $v_{\parallel}$ of the pellet, such that the LF\_HV\_Ne10 is comparable to a case with 222 m/s and 25 deg. shatter angle.}
\label{tab1}
\begin{tabular}{||l|c|c|r|c|c|r||}
Name           & Ne atoms           & D atoms            & Ne frac.    & Pre-shattered speed   & Shatter angle  & Frag. Num. \\
LF\_FV\_Ne0    & 0                  & $3.0\times10^{22}$ & 0\,\%         & 443 m/s & $12.5^\circ$     & 53    \\
LF\_FV\_Ne0.12 & $3.6\times10^{19}$ & $3.0\times10^{22}$ & 0.12\,\%      & 443 m/s & $12.5^\circ$     & 53    \\
LF\_FV\_Ne1    & $3.0\times10^{20}$ & $3.0\times10^{22}$ & 1\,\%         & 443 m/s & $12.5^\circ$     & 53    \\
LF\_FV\_Ne10   & $3.0\times10^{21}$ & $2.7\times10^{22}$ & 10\,\%        & 443 m/s & $12.5^\circ$     & 53    \\
MF\_FV\_Ne0.12 & $3.6\times10^{19}$ & $3.0\times10^{22}$ & 0.12\,\%      & 443 m/s & $20.0^\circ$     & 199   \\
SF\_FV\_Ne0.12 & $3.6\times10^{19}$ & $3.0\times10^{22}$ & 0.12\,\%      & 443 m/s & $25.0^\circ$     & 1105  \\
MF\_FV\_Ne10   & $3.0\times10^{21}$ & $2.7\times10^{22}$ & 10\,\%        & 443 m/s & $20.0^\circ$     & 199   \\
SF\_FV\_Ne10   & $3.0\times10^{21}$ & $2.7\times10^{22}$ & 10\,\%        & 443 m/s & $25.0^\circ$     & 1105  \\
LF\_HV\_Ne10*  & $3.0\times10^{21}$ & $2.7\times10^{22}$ & 10\,\%        & 222 m/s & $25.0^\circ$     & 53    \\
MF\_HV\_Ne10*  & $3.0\times10^{21}$ & $2.7\times10^{22}$ & 10\,\%        & 222 m/s & N/A           & 199   \\
\end{tabular}
\end{table*}

\begin{figure*}[htb!]
\centering
\includegraphics[width=.9\textwidth]{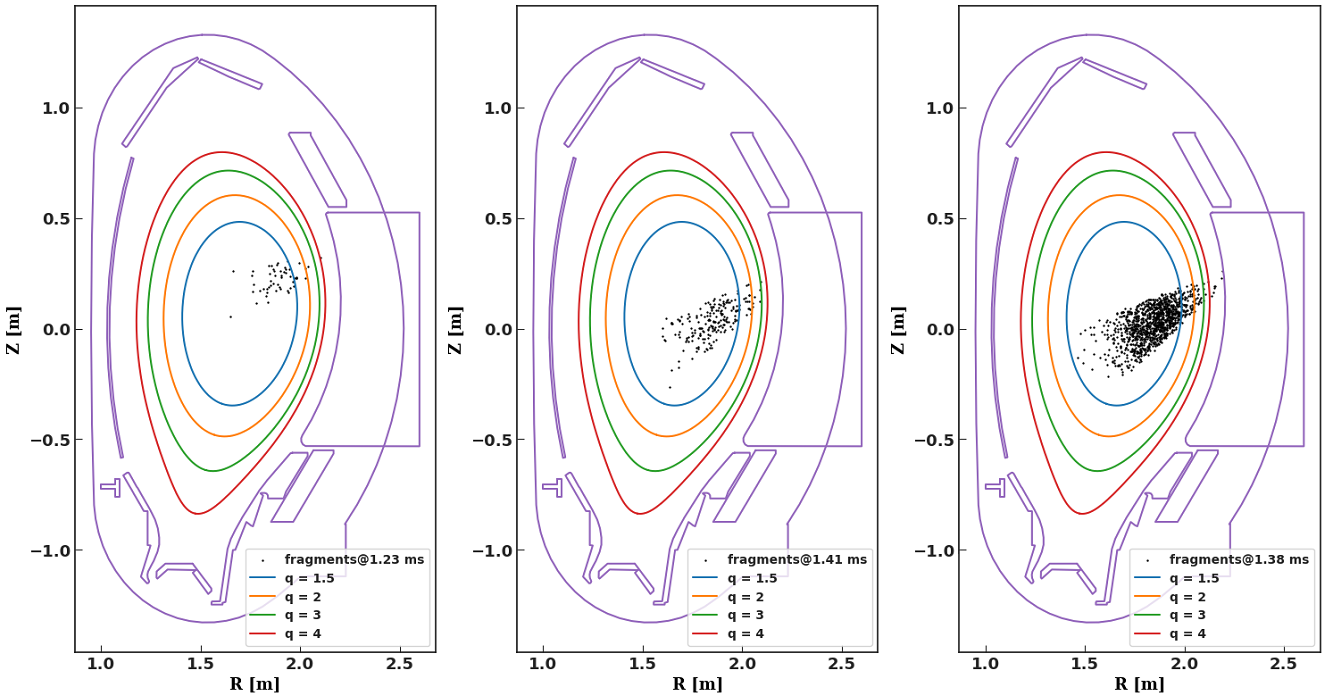}
\caption{Fragments plumes generated from the 443 m/s pellets. From left to right, the shattering angles are set as $12.5^\circ$ (GT3 in AUG), $20^\circ$ and $25^\circ$ (GT1 in AUG), respectively. The produced fragment numbers are accordingly 53, 199 and 1105. The contours indicate the locations of different equilibrium resonant surfaces.}
\label{fig_shards}
\end{figure*}

\begin{figure*}[htb!]
\centering
\includegraphics[width=.9\textwidth]{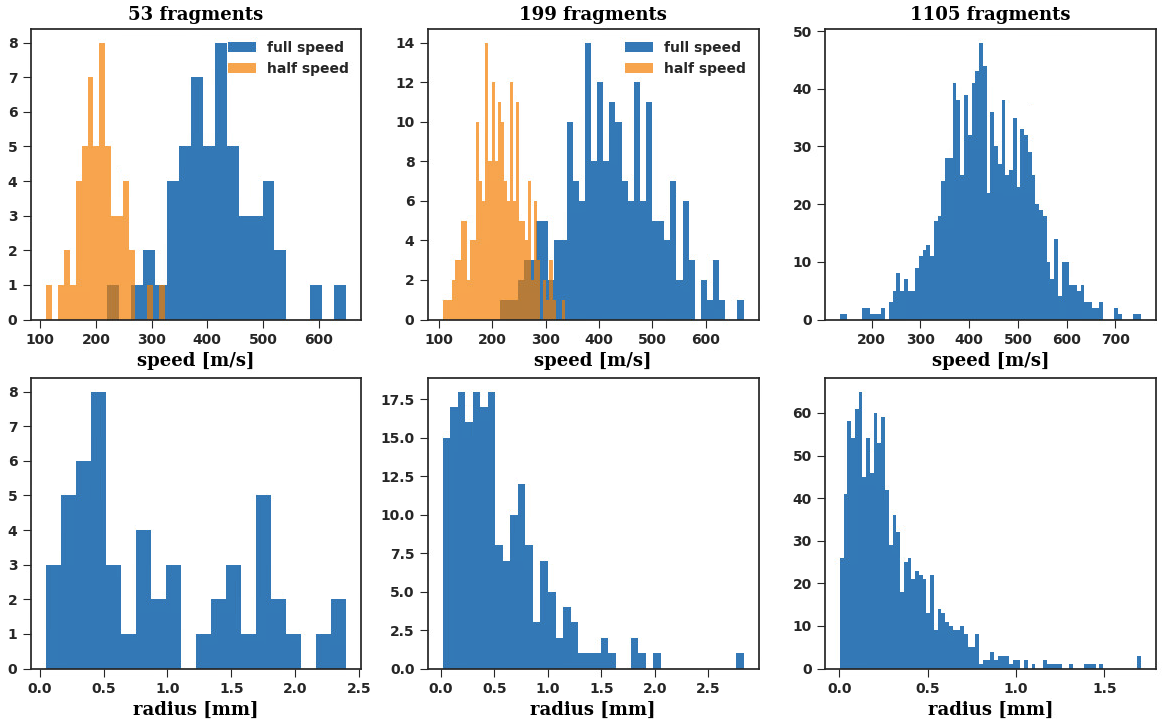}
\caption{Speed and radius distributions of the fragments. The left column illustrates the distributions for the 53 fragment cases (LF\_FV\_Ne0, LF\_FV\_Ne0.12, LF\_FV\_Ne1 and LF\_FV\_Ne10), the middle column for the 199 fragment cases (MF\_FV\_Ne0.12 and MF\_FV\_Ne10), and the right column for the 1105 fragment cases (SF\_FV\_Ne0.12 and SF\_FV\_Ne10). The two half speed cases (LF\_HV\_Ne10 and MF\_HV\_Ne10) are shown in orange.}
\label{fig_fragdist}
\end{figure*}

\begin{figure*}[htb!]
\centering
\includegraphics[width=.9\textwidth]{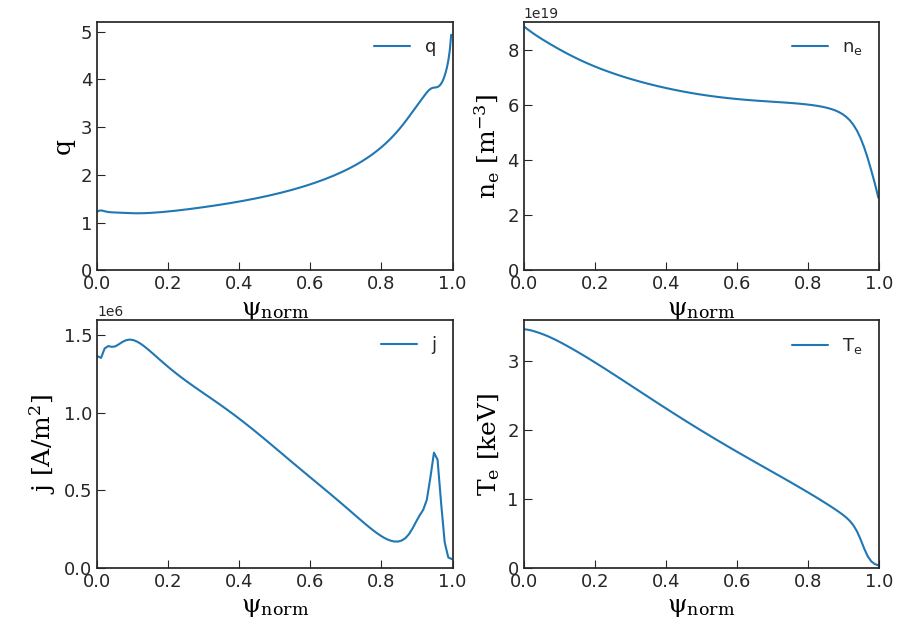}
\caption{Equilibrium profiles of safety factor $q$, electron density $n_e$, toroidal current density $j$ and electron temperature $T_e$.}
\label{fig_eq}
\end{figure*}

\begin{figure*}[htb!]
\centering
\includegraphics[width=.6\textwidth]{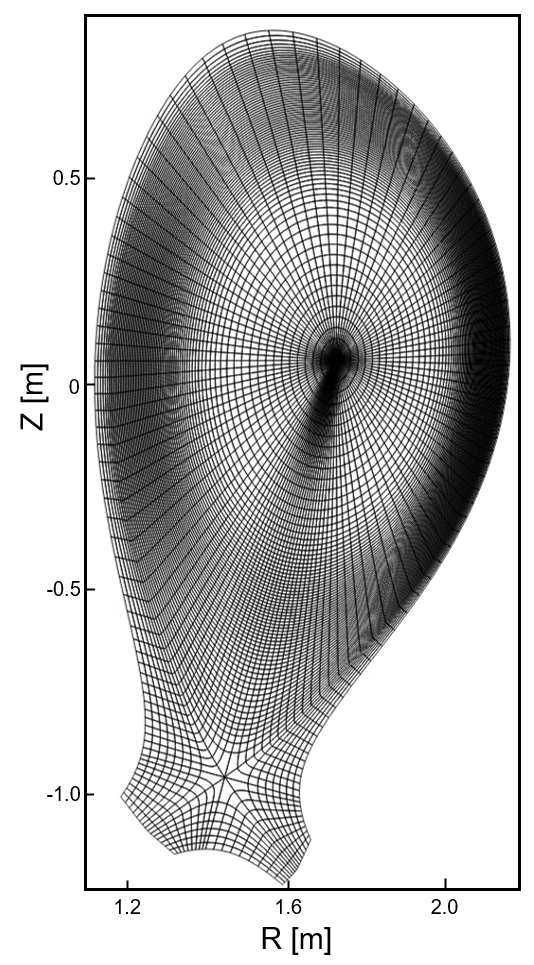}
\caption{Field aligned finite element grid including X-point used in this work. The grid resolution is 70 $\times$ 110 in the radial and poloidal direction, respectively.}
\label{fig_grid}
\end{figure*}

    \subsection{Fragment configuration}\label{shards}
        The fragments of the cryogenic pellet after shattering are assumed to be perfect spheres in the simulations. A rejection sampling method is applied to sequentially generate the radius of each fragment, presently based on Parks' fragmentation model~\cite{parks}. The sampling process is terminated once the accumulated volume of the fragments is equal to that of the injected pellet. The velocities of the fragments are generated based on the experimental observation that the mean fragment speed tends to lie between the pellet velocity parallel to the shatter plane $v_{para}$ and the injection speed of the pellet $v_{pellet}$~\cite{tobias}. For this reason, we assume the fragment velocity follows a Gaussian distribution with a mean $(v_{para}+v_{pellet})/2$ and standard deviation of 20\% of the mean. The directions of the fragment velocities are uniformly sampled on the unit sphere within the boundaries of a circular cone whose apex is located at the center of the sphere with a $20^\circ$ opening angle. The starting position of all fragments is the midpoint between the bend and the exit point of the shatter tube, and the average fragment direction is set parallel to the line defined by these two points and going towards the plasma.
        
        Three different shatter heads were installed on AUG during the 2022 experimental campaign~\cite{Dibon,HEINRICH2024114576}. To mimic the realistic fragment plume and compare the effects of different fragment sizes, three different fragment plumes are generated according to the above procedure, as shown in figure~\ref{fig_shards}. The pre-shattered speed of the pellet is set to 443~m/s for the base scenario, with a pellet length of 10~mm and a radius of 8~mm, matching the full-sized mean pellet length around 9.6~mm observed for the AUG system~\cite{HEINRICH2024114576}. By varying the shattering angle from $12.5^\circ$~-~$25^\circ$, different fragment sizes can be obtained while keeping a comparable penetration speed ($cos(12.5^\circ)=0.976$, $cos(25^\circ)=0.906$). The $25^\circ$ and $12.5^\circ$ cases correspond to the setup of guide tube 1 (GT1) and GT3 in AUG, respectively. Although it is best to control the variable when comparing the effects of fragment size, the differences in injector geometry and trajectory were preserved in order to accurately reflect the experimental conditions for each case. To investigate the effect of penetration speed, two additional fragment configurations were created by maintaining the same fragment size distributions as in the 443~m/s cases with 53 and 199 fragments, respectively, while manually reducing the velocities of the fragments to half of their original values. This is done to separate the variables to specifically study the effect of velocity. All the fragment configurations used in this study are listed in Tab.~\ref{tab1}, with their speed and radius distributions illustrated in figure~\ref{fig_fragdist}. Due to the high computational cost of 3D simulations, we were unable to perform scan over different random samplings of the fragment distributions within this work. However, complementary studies using the lower dimensional model DREAM have shown that different realizations of the same statistical distribution do not significantly alter the big picture, and only have a notable impact on the disruption dynamics for trace amounts of injected neon~\cite{peter2024}. In the simulation, each fragment has its own velocity and radius, and follows an individual trajectory originating from its initial location. The velocity vector is kept constant while the radius is updated according to the ablation of the fragment on its course through the plasma domain (rocket effect~\cite{rocketprl} not taken into account).

\begin{figure}[htb!]
\centering
\includegraphics[width=.45\textwidth]{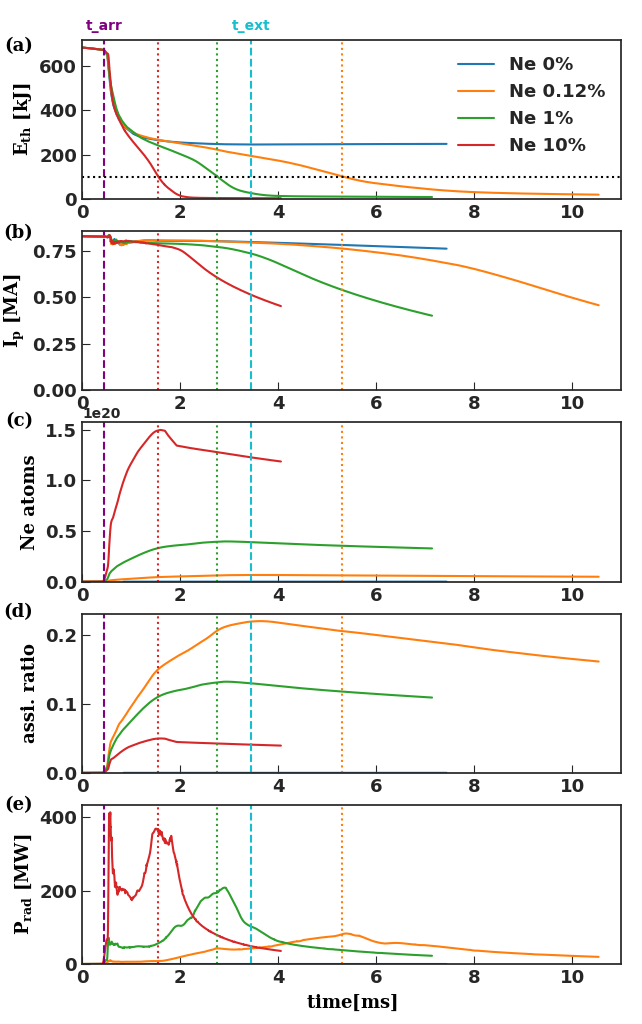}
\caption{Temporal evolution of (a) total thermal energy, (b) plasma current, (c) total assimilated neon quantity, (d) ratio of assimilated neon to total injected neon and (e) radiation power for the 53 fragment cases with different neon fraction. The time of fragment arrival $t_{arr}$ and exit $t_{ext}$ are marked by the purple and cyan vertical dashed lines, respectively. The times at the end of TQ are annotated by the vertical dotted lines.}
\label{fig_neonscan}
\end{figure}

\begin{figure*}[htb!]
\centering
\includegraphics[width=1.\textwidth]{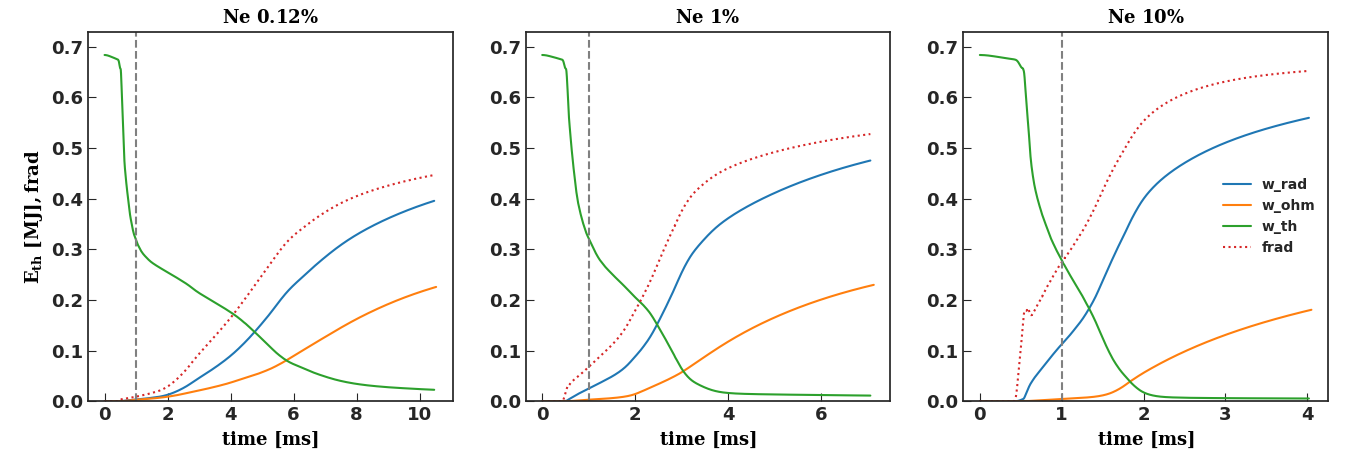}
\caption{Temporal evolution of total radiated energy, ohmic heating energy and total thermal energy for the 53 fragment cases with different neon fraction. The dotted lines show the real-time radiation fraction $frad(t)=W_{rad}(t)/[W_{th}(0)-W_{th}(t)+W_{ohm}(t)]$. The vertical dashed lines indicate the time point at t = 1~ms, roughly the dividing line between the first and second stage of cooling.}
\label{fig_wrad}
\end{figure*}

\begin{figure*}[htb!]
\centering
\includegraphics[width=1.\textwidth]{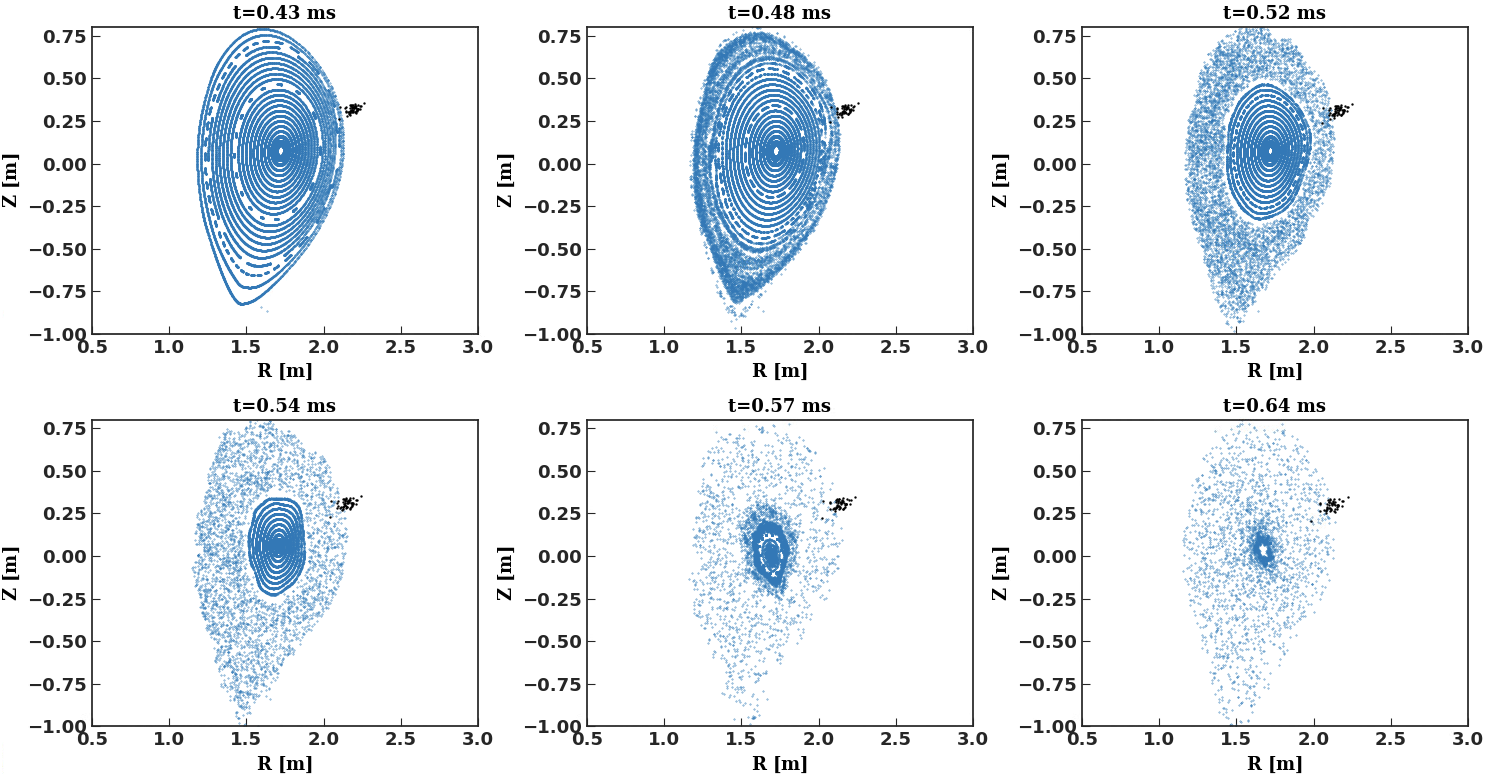}
\caption{Poincar\'{e} plots of magnetic fields for the pure deuterium injection with 53 fragments. The black markers indicate the real-time location of the fragments that are not completely ablated.}
\label{fig_poincareD}
\end{figure*}

\begin{figure*}[htb!]
\centering
\includegraphics[width=.8\textwidth]{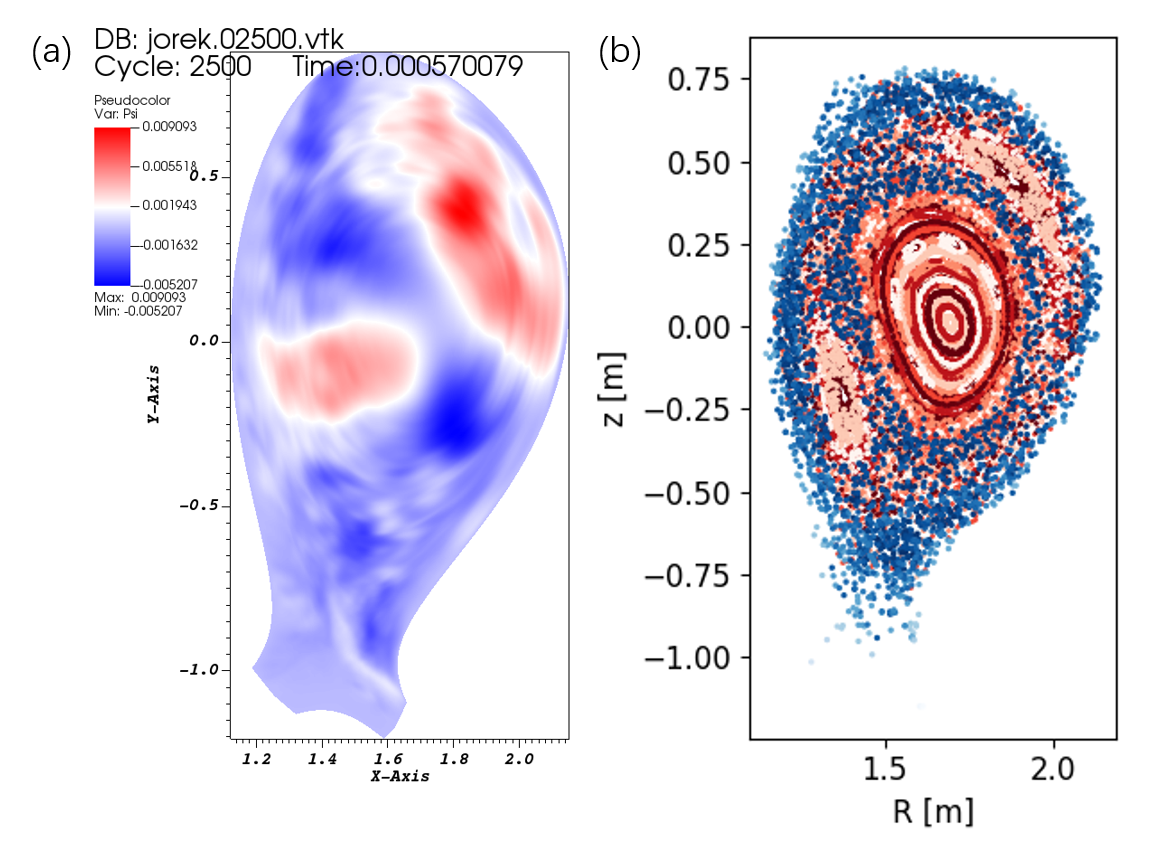}
\caption{(a)Perturbed poloidal magnetic flux $\delta \psi$ and (b)Poincar\'{e} plot for the same case as figure~\ref{fig_poincareD} at 0.57 ms. Only n=0 and n=1 toroidal harmonics of field lines are traced here to visualize the underlying island structures behind the stochastic fields.}
\label{fig_21mode}
\end{figure*}

\begin{figure*}[htb!]
\centering
\includegraphics[width=1.\textwidth]{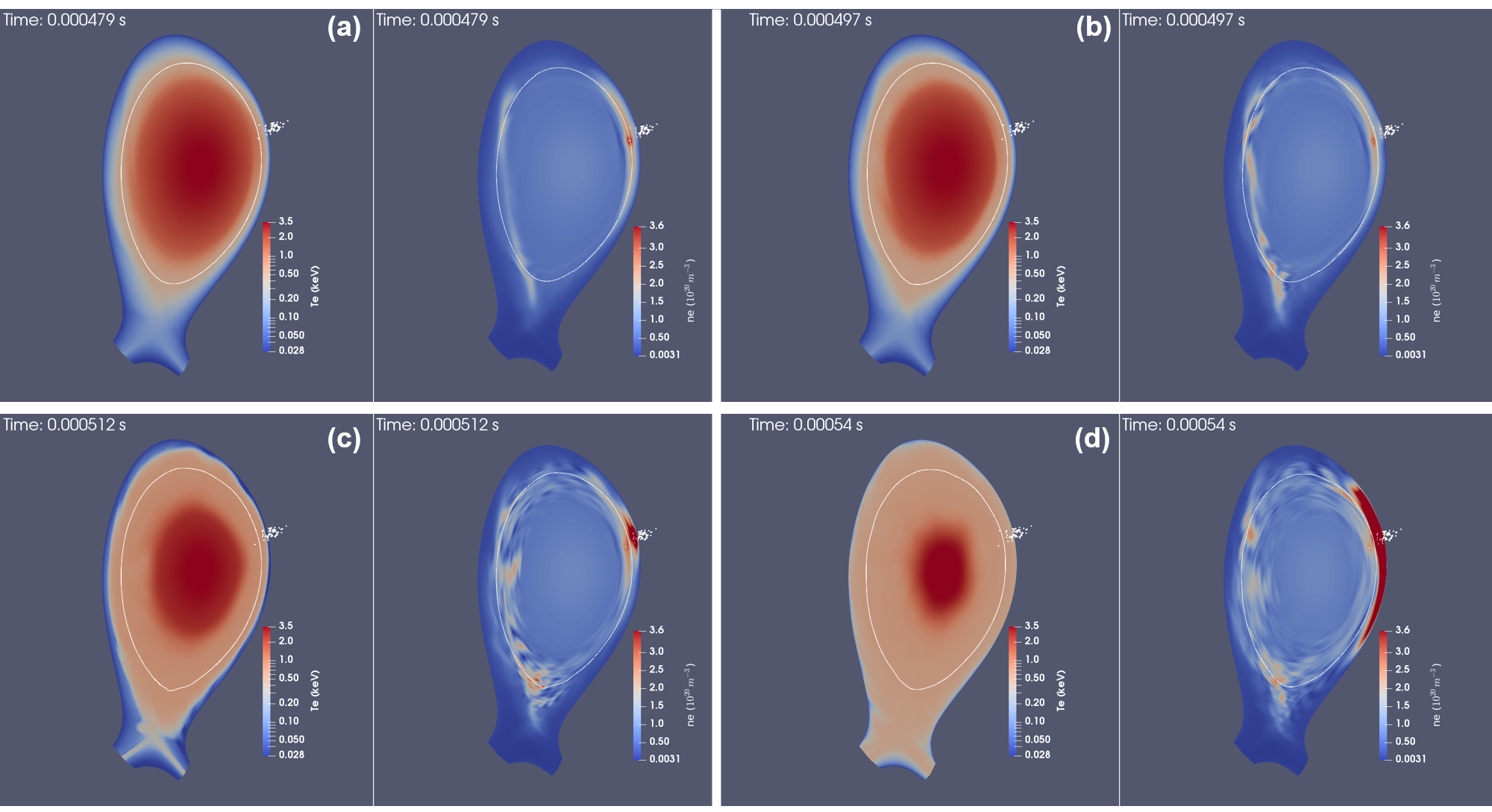}
\caption{Electron temperature and density at different times (a) 0.479~ms (b) 0.497~ms (c) 0.512~ms (d) 0.540~ms during the inital thermal energy drop phase for the pure deuterium case with 53 fragments. The white markers indicate the real-time location of the injected pellet fragments. The white contour illustrates the q~=~3 resonant surface.}
\label{fig_field_pureD}
\end{figure*}

\begin{figure*}[htb!]
\centering
\includegraphics[width=.85\textwidth]{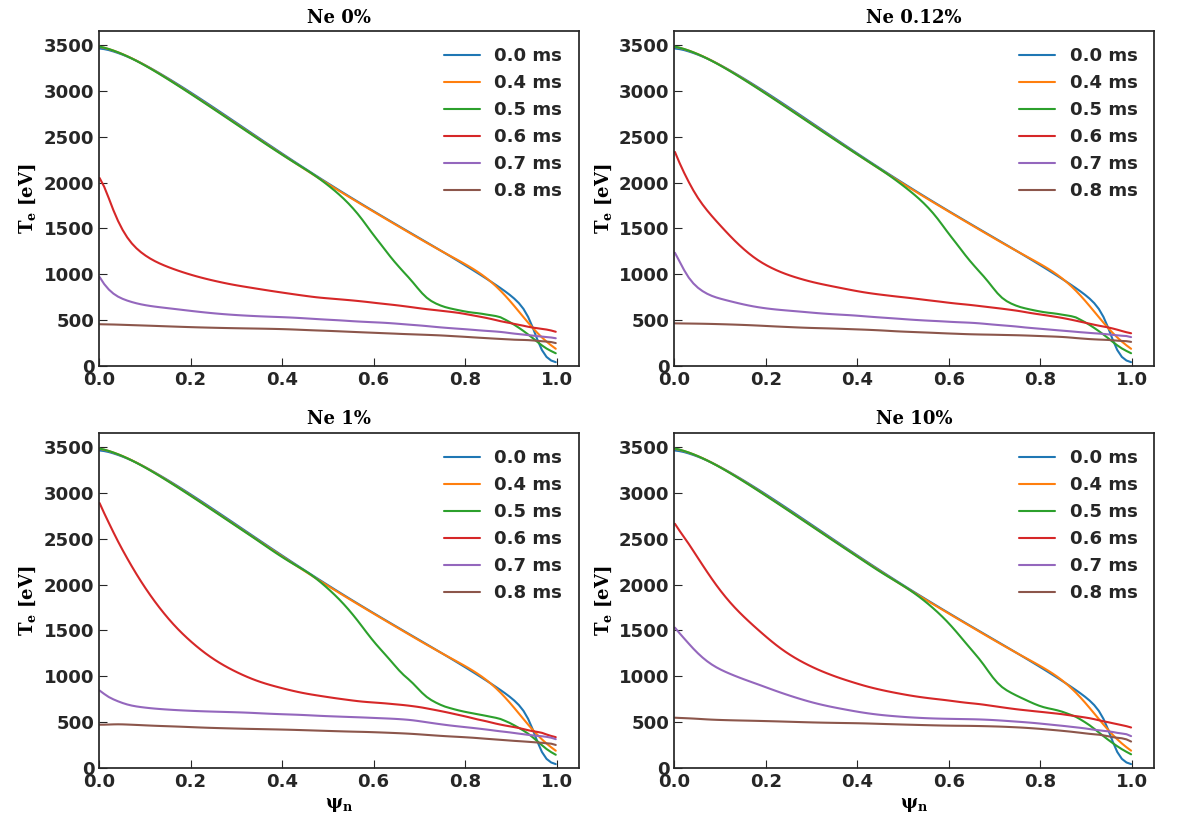}
\caption{Flux-averaged temperature profile evolution for different neon fractions during the first stage cooling.}
\label{fig_teprof_ne}
\end{figure*}

\begin{figure*}[htb!]
\centering
\includegraphics[width=.8\textwidth]{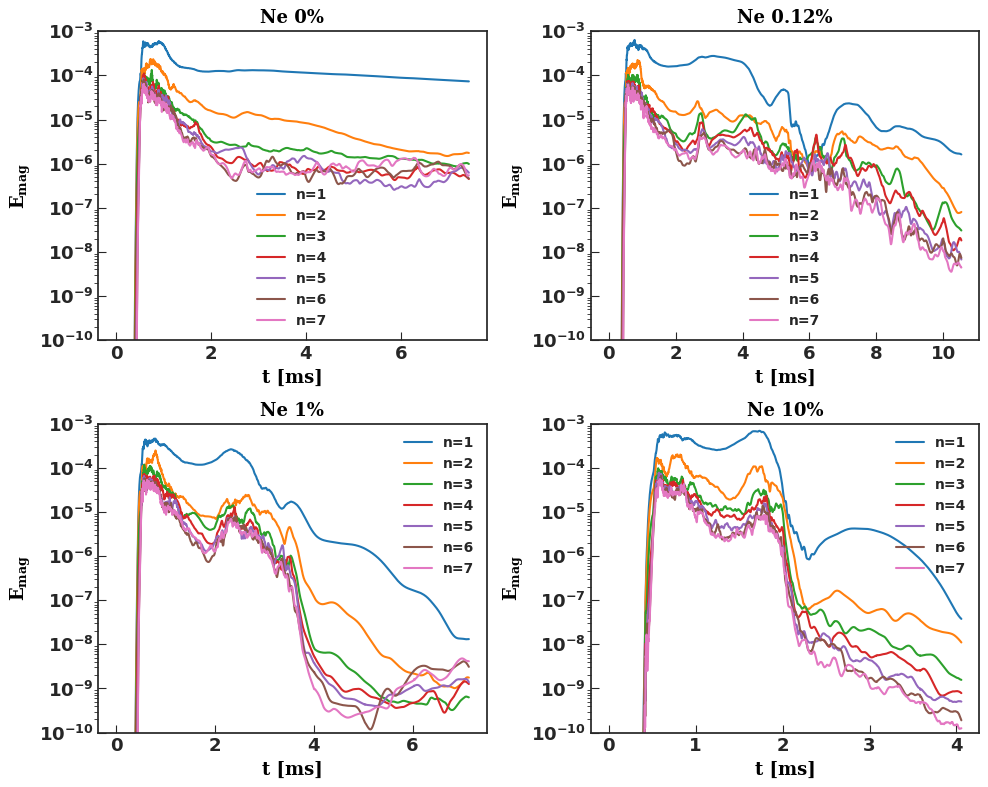}
\caption{Magnetic energy spectrum evolution for the 53 fragment cases with different neon fraction.}
\label{fig_mag_ne}
\end{figure*}

\begin{figure*}[htb!]
\centering
\includegraphics[width=1.\textwidth]{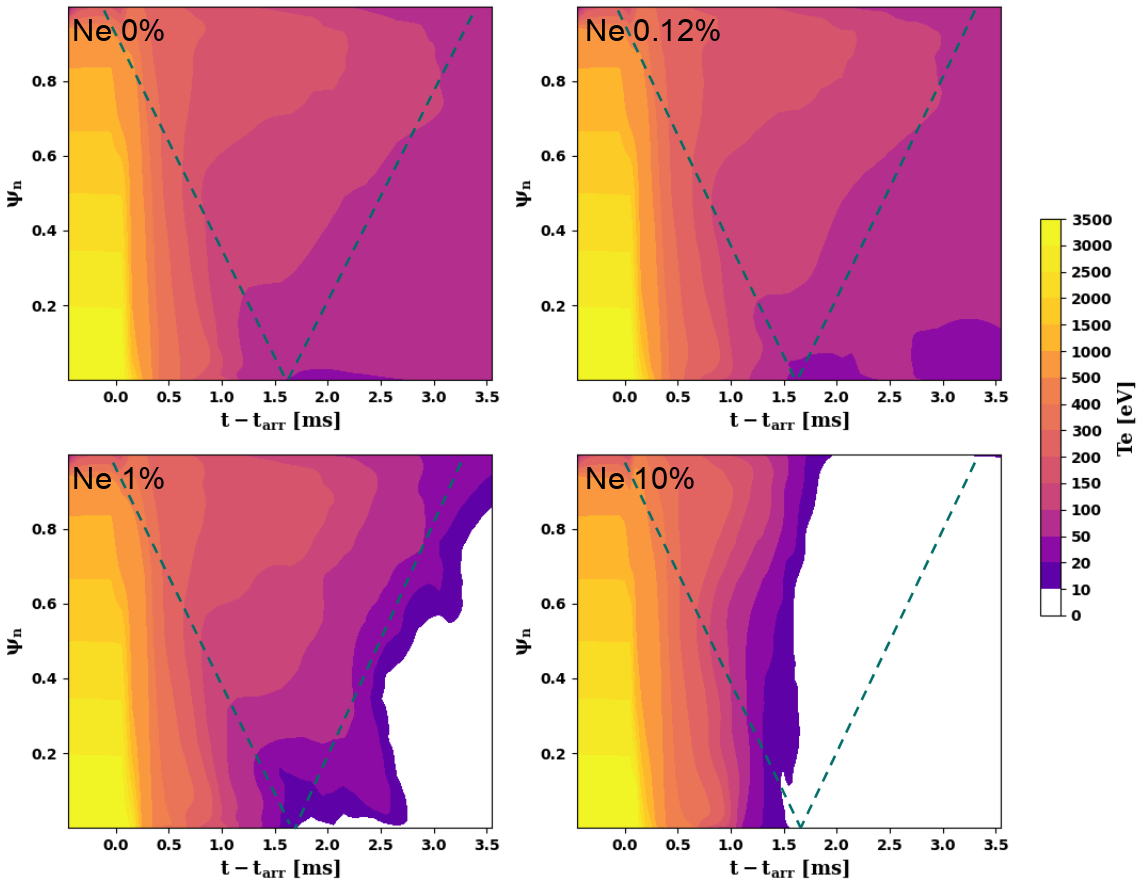}
\caption{Flux-averaged temperature profile evolution for the 53 fragment cases with different neon fraction. The dashed lines indicate the estimated trajectory of the bulk fragments. For all these cases, temperature profiles are drastically degraded following the fragments arrival (0 - 0.5~ms). As the fragments penetrate inward, a localized low-temperature region is created and move inward along with the fragments (0.5 - 1.5~ms). As for the neon mixed injections, the core temperature suddenly drops to below 10 eV following the 1/1 MHD activity, and then initiating the inside-out cooling. }
\label{fig_tecont_ne}
\end{figure*}
        
\section{Numerical results}\label{sec3}
    \subsection{Simulation setup}\label{setup}
        The equilibrium profiles, shown in figure~\ref{fig_eq}, are based on the reconstruction of AUG H-mode discharge \#40355 with CLISTE~\cite{cliste}. The toroidal field $B_t$ is 1.8~T, and the plasma current $I_p$ is 0.8~MA. A Spitzer-like resistivity $\eta$ is adopted in the simulation: 
        \begin{equation}
        \eta= \eta_0 T_e^{-3 / 2} \cdot Z_{\mathrm{eff}} \frac{1+1.198 Z_{\mathrm{eff}}+0.222 Z_{\mathrm{eff }}^2}{1+2.966 Z_{\mathrm{eff}}+0.753 Z_{\mathrm{eff}}^2},
        \end{equation}
        where $Z_{\mathrm{eff}}$ is the effective charge,
        \begin{equation}
        Z_{\mathrm{eff}} \equiv \frac{\sum_i n_i Z_i^2}{\sum_i n_i Z_i}.
        \end{equation}
        In order to prevent infinite $\eta$ values at extremely low temperature, a temperature threshold $T_{min}$~=~1~eV is used to cut off the temperature dependency, i.e., when the temperature drops below this value, $\eta$ will not further increase. In the same vein, an upper limit of temperature dependency is applied at $T_{max}$~=~1.35~keV, corresponding to a $\eta$ floor value of $2~\times~10^{-8}~\Omega$m to simplify the simulation numerically. The perpendicular viscosity $\mu$ and parallel viscosity $\mu_{\parallel}$ are both set as $4.85\times10^{-7} ~\mathrm{kg~m^{-1}s^{-1}}$. The corresponding kinematic viscosity in the center $\nu_0=\mu/\rho_0$ is $1.6\ \mathrm{m^2/s}$. The perpendicular thermal diffusivity is set as $\chi_\perp = 1\ \mathrm{m^2/s}$. Spitzer-Haerm parallel thermal diffusivity, 
        \begin{equation}
        \chi_{\|, S H}=3.6 \cdot 10^{29} \frac{T_e[\mathrm{keV}]^{5 / 2}}{n_{e}\left[m^{-3}\right]} m^2 / \mathrm{s},
        \end{equation}
        is adopted in the parallel direction. A floor value at 50~eV is applied in the simulation to help deal with low temperature regions such as temperature holes. The isotropic particle diffusivity for main ions $D$ is set to be 1.6 $\mathrm{m^2/s}$. The impurity diffusivity is set to be the same with the main ions, i.e. $D_{imp}=1.6\ \mathrm{m^2/s}$. No background impurity and no initial magnetic perturbation are included in the simulations to keep the focus on the injected material. 
        
        The simulations utilize a flux-aligned grid, as shown in figure~\ref{fig_grid}, with a resolution of 70 $\times$ 110 in the radial and poloidal direction, respectively. After rigorous convergence tests on number of toroidal Fourier modes, 8 toroidal harmonics (n = 0...7) are chosen for the production simulations. Since the simulations are aimed for late TQ to early CQ, an ideal conducting wall is assumed. The nominal time step used in our simulations is approximately $6.1\times10^{-8}$ s, which corresponds to 0.1 times the normalized JOREK time unit $t_{norm}$, during the CQ phase, the time step can be gradually increased to 1~-~2 times $t_{norm}$.

        Before launching our production simulations, we conducted a preliminary assessment to ensure that the diffusive evolution of the plasma equilibrium would not interfere with the results. Specifically, we examined the evolution of the equilibrium profiles over a period of 1.2~ms without any injection. During this interval, the equilibrium changes only slightly, with no significant alteration in the global profiles. The SPI is launched at the very beginning in our simulations. Once a fragment enters the plasma domain, the ablation rate for a mixed neon and deuterium fragment is calculated based on an improved version~\cite{parksabl} of the neutral gas shielding (NGS) model~\cite{parks1978,Gal_2008}. The ablation rate of each fragment is obtained by 
        \begin{eqnarray}
            G[g/s] = &\ [27.08 + \mathrm{tan}(1.49X)] \left(\frac{T_e[\mathrm{eV}]}{2000}\right)^{5/3}\nonumber\\ &\times \left(\frac{r_p[\mathrm{cm}]}{0.2}\right)^{4/3} \left(n_e[10^{14}\mathrm{cm}^{-3}]\right)^{1/3},
        \end{eqnarray}
        where $X=N_{D_2}/(N_{D_2}+N_{Ne})$ is the molecular mixture ratio of deuterium and $r_p$ is the radius of the fragment.
        The ablated material is assumed to be a Gaussian shape source term centered at the fragment location, with the following form:
        \begin{eqnarray}
            S_n \propto &\exp \left(-\frac{\left(R-R_{\mathrm{s}}\right)^2+\left(Z-Z_{\mathrm{s}}\right)^2}{\Delta r_{\mathrm{NG}}^2}\right) \quad \nonumber\\
                        &\times \exp \left(-\left(\frac{\phi-\phi_{\mathrm{s}}}{\Delta \phi_{\mathrm{NG}}}\right)^2\right).\label{eqsrs}
        \end{eqnarray}    
        In equation (\ref{eqsrs}), $R_{\mathrm{s}}, Z_{\mathrm{s}}$ and $\phi_{\mathrm{s}}$ are the real-time location of one specific fragment. $\Delta r_{\mathrm{NG}}$ is the radius of the neutral gas cloud in the poloidal plane, which is set to 0.03~m. The poloidal grid spacing in our setup is approximately 0.03~m, which already represents a very fine mesh given the global scale of the simulations. Using a narrower deposition width would not be numerically meaningful in this context, as it would fall below the grid resolution and lead to under-resolved or numerically unstable behavior. $\Delta \phi_{\mathrm{NG}}$ is the toroidal extent of the cloud, which is set to 0.8 rad in our simulation, due to the limitation of the toroidal grid resolution.

    \subsection{Neon content scan}\label{neonscan}   
        To investigate the effect of the neon concentration, a scan in the neon content is performed at fixed number of fragments (LF\_FV\_Ne0, LF\_FV\_Ne0.12, LF\_FV\_Ne1 and LF\_FV\_Ne10 in Tab.~\ref{tab1}). To isolate the effects of the different neon fractions, we assume an identical size and velocity of the fragments in the scan. The total injected neon and deuterium atoms are listed in Tab.~\ref{tab1}. 
    
        Figure~\ref{fig_neonscan} shows the evolution of global quantities over time after the injection. From the thermal energy evolution in figure~\ref{fig_neonscan} (a), it can be seen that a very rapid drop of thermal energy is triggered upon the fragments arriving at the plasma domain at about 0.45 ms, regardless of the neon fraction. The sudden drop of about 50\% of the thermal energy is mainly due to the convective and conductive transport caused by the magnetic field line stochastization at the edge. For the pure deuterium case (0\% neon), the plasma enters a prolonged non-disruptive phase following the initial drop in thermal energy. In contrast, for neon-doped pellet cases, the thermal energy is gradually depleted after the initial drop. The pre-TQ+TQ duration (period from fragments arrival to the end of TQ) increases as the neon fraction decreases. The time of fragment arrival and exit are marked by the purple and cyan vertical dashed lines, respectively. The times at the end of TQ are annotated by the colorful vertical dashed lines in the plot when the total thermal energy drops to about 100 kJ, which are in line with the the radiation peaks in figure~\ref{fig_neonscan} (e). 
        
        In figure~\ref{fig_neonscan} (c), the total assimilated neon quantity is plotted, while the assimilation ratio of impurity neon, calculated by the ratio of assimilated neon to the total injected neon content, is shown in figure~\ref{fig_neonscan} (d). The 10\% neon pellet has the largest total assimilated neon quantity, whereas its assimilation ratio is the lowest comparing to other neon fraction cases. This is a self-regulating process that, with a lower neon fraction, the electron temperature decreases more gradually, resulting in an extended ablation period and consequently a higher assimilation ratio, which is consistent with the INDEX results in Ref.~\cite{ansh}. It should be noted that for the 10\% and 1\% neon cases, ablation stops at approximately 2 and 3 ms, respectively, when the total thermal energy drops to nearly 100 kJ at the end of TQ. However, for the 0.12\% neon case, ablation saturates at around 3.5 ms as marked by the cyan vertical dashed line, while the plasma temperature remains sufficiently high for fragment ablation. This occurs because, at that moment, the majority of the fragments have traversed the plasma and start to exit the plasma domain. It is worth noting that even for a pellet with a low neon fraction of 0.12\%, the ablation ratio remains limited to approximately 20\%, indicating that a significant amount of material still remains unablated in the simulation.
        
        Figure~\ref{fig_neonscan} (e) shows the evolution of radiation power versus time, where double radiation peaks are observed for the higher neon fraction cases (1\% and 10\%). Analysis of the radiation distribution reveals that the first radiation peak originates from the edge near the injection location, while the second peak is associated with the plasma core. The first radiation peak is due to the initial ablation of the fragments, while the second peak is related to the 1/1 MHD mode during core collapse.

        To understand the two-stage cooling observed in our simulations, figure~\ref{fig_wrad} displays the evolution of total radiated energy, ohmic heating energy, total thermal energy and radiation fraction for the cases with 0.12\%, 1\% and 10\% neon. It shows that, before 1~ms (indicated by the gray dashed line), the radiation fraction is very limited. In the cases with 0.12\%, 1\%, and 10\% neon, the radiation fractions at that time are approximately 0\%, 6\%, and 27\%, respectively. During the second stage of thermal energy drop (after 1 ms), energy loss is dominated by radiative cooling, with the radiation fraction increasing to 80\%, 93\% and 97\%, respectively, as the neon content increases. Here, the radiation fraction is calculated using the formula $\Delta W_{rad}/(\Delta W_{th}+\Delta W_{ohm})$, where $\Delta W_{rad}$ represents the change in radiated energy, $\Delta W_{th}$ is the change in thermal energy, and $\Delta W_{ohm}$ refers to the change in Ohmic heating energy. The calculation is separated into two distinct time windows of the thermal energy drop. Due to the absence of background impurity and lack of impurity radiation, the second phase may not occur at all in pure deuterium injection, like also observed in many non-disruptive shots in experiments~\cite{paul2024arxiv}.
        
        To study the first stage of thermal energy loss in more depth, the Poincar\'e plots of magnetic fields for the pure deuterium injection case are plotted in figure~\ref{fig_poincareD}. The plot shows that right after the vanguard fragments reach the plasma edge (0.48 - 0.52 ms), the magnetic field becomes stochastic near the fragment locations, due to the destabilization by helical cooling. Since the injection targets a H-mode plasma, the large current density gradient and pressure gradient in the pedestal are prone to MHD instabilities, triggering magnetic reconnection and leading to the stochastization. Due to the steep pressure gradient at the pedestal, consequently, considerable energy loss occurs during the early injection phase once confinement at edge is broken. Afterwards, when fragments reach the q = 2 surface at around t = 0.57 ms, a large 2/1 island is induced and triggers the global reconnection event, resulting in full stochastisation of the flux surface. This large 2/1 island can be seen by producing Poincar\'e plots, which include only the n=0 and n=1 toroidal harmonics of the magnetic field, as shown in figure~\ref{fig_21mode} (b). In figure~\ref{fig_21mode} (a), the perturbed poloidal magnetic flux $\delta \psi$ at the same time is plotted, revealing a 2/1 tearing mode structure.
        
        The evolution of electron temperature and density the during first stage of thermal energy loss are shown in figure~\ref{fig_field_pureD}, and the flux-averaged electron temperature profiles are plotted in figure~\ref{fig_teprof_ne}. Upon the vanguard fragments arriving at the q~=~3 surface at 0.479~ms, the 3/1 magnetic islands are induced, as the m~=~3 structure of the ablated electron density becomes clearly visible in figure~\ref{fig_field_pureD} (a). The pedestal is subsequently degraded, and the temperature becomes flattened near the q~=~3 resonant surface at around t~=~0.5~ms, as shown in figure~\ref{fig_field_pureD} (b) and in the flux-averaged profile in figure~\ref{fig_teprof_ne}. Following the pedestal collapse, significant heat flux propagates outward to the edge, causing a sudden ablation of the subsequent fragments. This is illustrated in figure~\ref{fig_field_pureD}(c) and (d), where considerable thermal energy is expelled from the edge. Figure~\ref{fig_teprof_ne} also indicates that the majority of thermal energy loss during the first stage occurs between 0.5 and 0.6~ms. The evolution of the temperature profile is similar across different neon fractions, due to the absence of significant radiative cooling at this stage. 
        
        During the first stage of the cooling, the majority of the fragments are located at the plasma edge. Despite significant degradation, the plasma temperature is still warm enough ($\approx$ 500 eV) for ablation to continue. With impurity radiation coming into effect, the second stage of cooling diverges depending on the neon fraction. The magnetic energy spectrum is compared in figure~\ref{fig_mag_ne}, for the four cases. In the pure deuterium case, only a single MHD peak is observed during the initial injection stage, corresponding to the rapid drop in thermal energy at the beginning. For neon-doped injections, in addition to the first MHD peak observed at the early stage, a later MHD peak appears at the end of the TQ phase, aligning with the radiation peaks. The second MHD peak corresponds to a 1/1 mode activity, which leads to the core collapse. In cases with a higher neon fraction, when the fragments reach the core, the core collapse is triggered shortly due to stronger MHD response. With low neon concentration, the radiative cooling is less intense, leading to milder MHD activity. As a result, the plasma remains more stable for a longer period, and the core collapse occurs after the fragments have exited the plasma domain.

        This slow TQ process can also be seen in figure~\ref{fig_tecont_ne}, where the temporal evolution of the flux-averaged temperature profiles are compared for different neon fractions. For all these cases, after the initial degradation of temperature, an inside-out cooling pattern is observed. As the fragments penetrate inward, they create a localized low-temperature region that moves inward along with the fragments, while a temperature layer of approximately 100 eV remains at the edge. This inside-out cooling does not occur at higher fragment numbers or lower penetration speed, as discussed later. For pure deuterium injection, due to lack of radiation, the thermal energy is maintained and the temperature profile gradually relaxes over time. In contrast, with neon mixed injections, the core temperature suddenly drops to below 10 eV following the 1/1 MHD activity, and then initiating the inside-out cooling. 

    \subsection{Influence of fragment size}\label{size}
        To understand the influence of fragment size on the disruption mitigation effectiveness, simulations with different fragment sizes are conducted with a neon fraction of 0.12\% and 10\%, corresponding to the cases MF\_FV\_Ne0.12, SF\_FV\_Ne0.12, MF\_FV\_Ne10 and SF\_FV\_Ne10 from Tab.~\ref{tab1}. 

\begin{figure}[htb!]
\centering
\includegraphics[width=.45\textwidth]{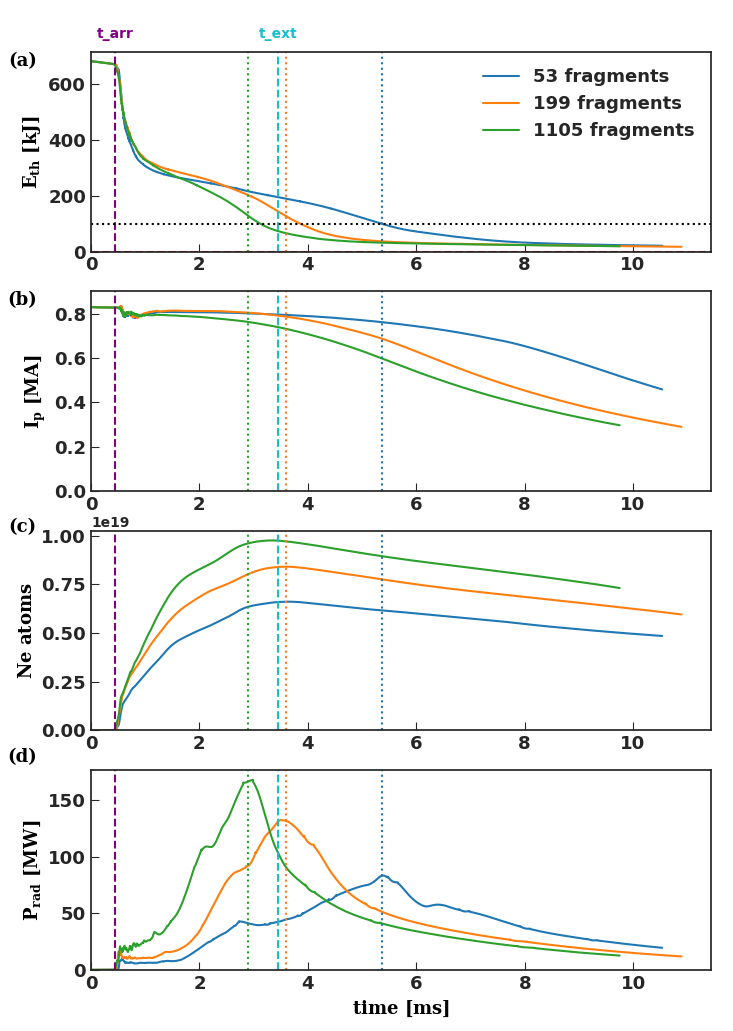}
\caption{Temporal evolution of (a) total thermal energy, (b) plasma current, (c) total assimilated neon quantity and (d) radiation power for different fragment numbers for fixed total volume/material with 0.12\% neon. The time of fragment arrival $t_{arr}$ and exit $t_{ext}$ at/from the plasma domain are marked by the purple and cyan vertical dashed lines, respectively. The times at the end of TQ are annotated by the vertical dotted lines.}
\label{fig_size012}
\end{figure}

\begin{figure}[htb!]
\centering
\includegraphics[width=.45\textwidth]{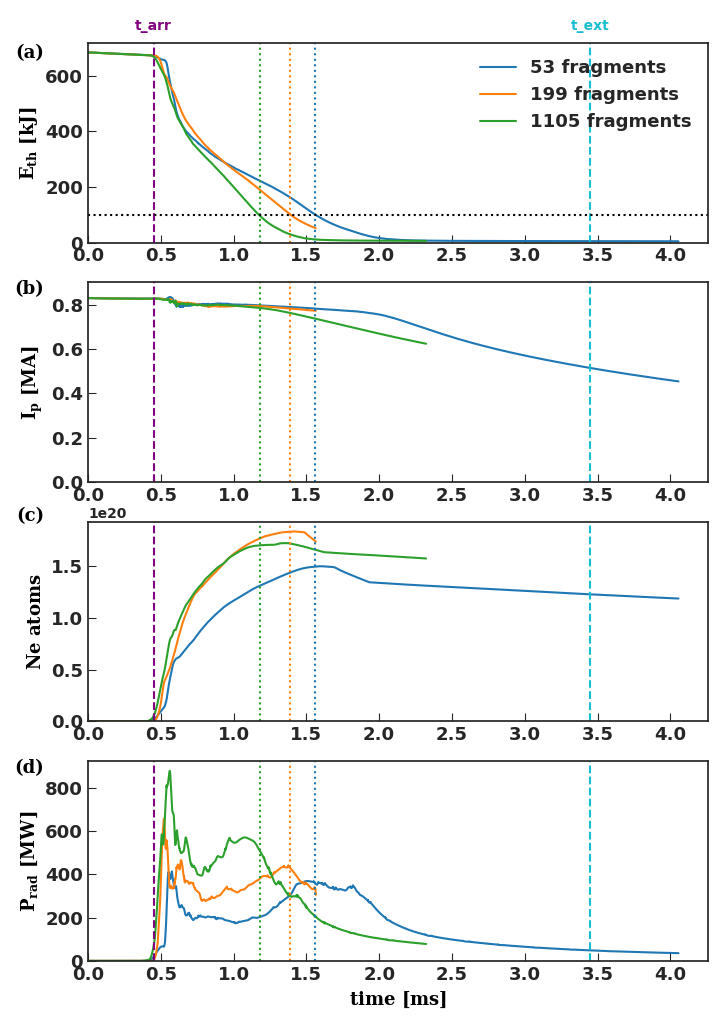}
\caption{Temporal evolution of (a) total thermal energy, (b) plasma current, (c) total assimilated neon quantity and (d) radiation power for different fragment numbers for fixed total volume/material with 10\% neon. The time of fragment arrival $t_{arr}$ and exit $t_{ext}$ at/from the plasma domain are marked by the purple and cyan vertical dashed lines, respectively. The times at the end of TQ are annotated by the vertical dotted lines.}
\label{fig_size10}
\end{figure}

\begin{figure}[htb!]
\centering
\includegraphics[width=.45\textwidth]{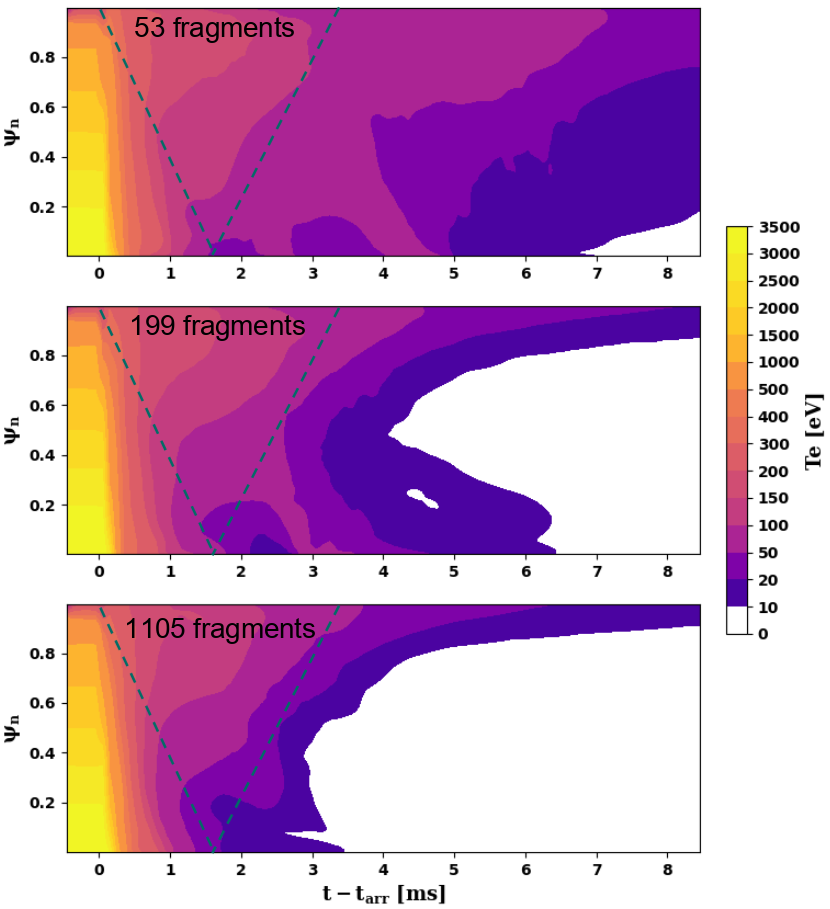}
\caption{Flux-averaged temperature profile evolution for different fragment numbers with 0.12\% neon. The dashed lines indicate the estimated trajectory of the bulk fragments.}
\label{fig_tecont_size012}
\end{figure}

\begin{figure}[htb!]
\centering
\includegraphics[width=.45\textwidth]{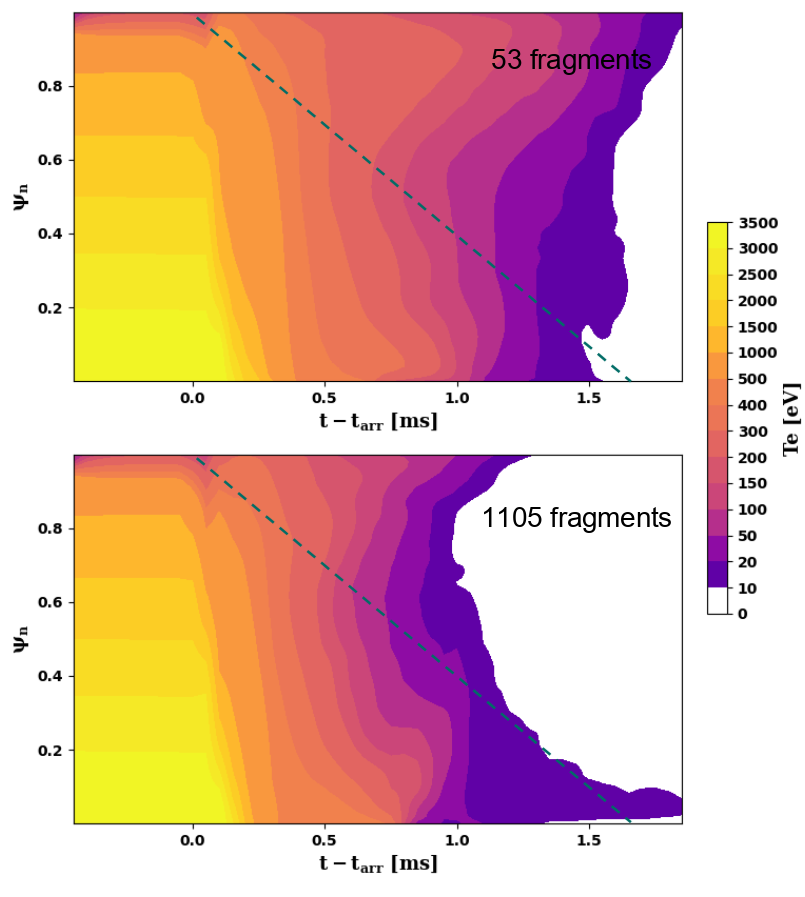}
\caption{Flux-averaged temperature profile evolution for different fragment numbers with 10\% neon. The dashed lines indicate the estimated trajectory of the bulk fragments.}
\label{fig_tecont_size10}
\end{figure}

\begin{figure*}[htb!]
\centering
\includegraphics[width=1.\textwidth]{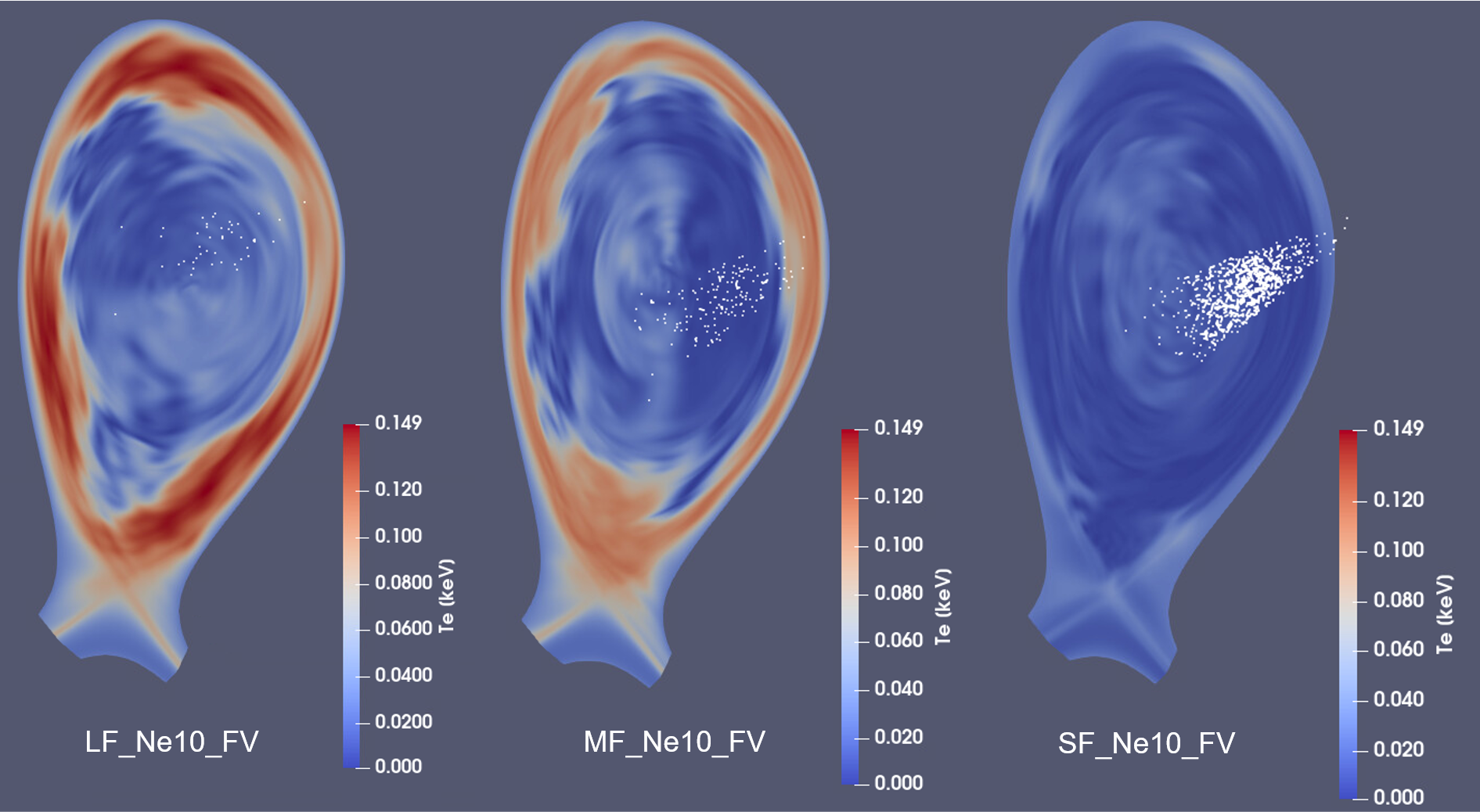}
\caption{ Electron temperature distribution near the end of the TQ for cases with varying numbers of fragments. As the number of fragments increases, the temperature belt around 100~eV near the edge is progressively eliminated, indicating more effective edge cooling. This enhanced cooling is attributed to the more elongated and sustained plume generated by smaller fragments, which improves assimilation during the initial injection stage.}
\label{fig_tebelt}
\end{figure*}%
        
        Figure~\ref{fig_size012} shows the overview plots for different fragment numbers in the 0.12\% neon scenario, where the total injected material is identical across cases. In figure~\ref{fig_size012} (a), a trend is observed that a larger number of smaller fragments can lead to a shorter TQ duration as indicated by the thermal energy evolution. This agrees with the lower dimensional results we previously obtained between JOREK and INDEX~\cite{Matsuyama_2022}. The larger total surface area for smaller fragments contributes to stronger assimilation until the majority of the fragments exit the plasma at around 3.5 ms. Since the end of ablation is not determined by a full thermal collapse here, but by the leaving of the fragments, the ablation for all cases saturates at almost the same time, which can be clearly seen in figure~\ref{fig_size012} (c). As the ablation duration is not constrained by the low temperature at TQ, smaller fragments achieve higher assimilation compared to larger ones. Owing to the larger assimilation, the core collapse occurs earlier with smaller fragments and stronger radiation power is observed at the core collapse phase, as shown in figure~\ref{fig_size012} (d). 
        
        Our numerical results deviate from the experimental findings reported by Jachmich~\cite{stefanEPS24}, where larger fragments with low neon fraction exhibit better material assimilation. Rocket effect and plasmoid drift are promising candidates for explaining this discrepancy. 
        Smaller fragments are more susceptible to the rocket effect due to smaller mass, which accelerates the fragments and deflects their trajectories. When multiple small fragments ablate simultaneously in a localized region, their combined neutral clouds can generate a larger local pressure peaking, and hence stronger plasmoid drift. In experiments~\cite{stefanEPS24}, smaller fragments tend to have longer pre-TQ duration yet achieve lower assimilation. Simulations using the NGS model indicate that smaller fragments ablate at a higher rate; if their total assimilation remains lower despite longer ablation durations, it is very likely due to the plasmoid drift and the rocket effect expelling the material from the plasma for low neon fraction injections.
        
        Figure~\ref{fig_size10} shows the same data, but for the 10\% neon cases with different fragment numbers. It can be observed that a larger number of smaller fragments leads to quicker assimilation due to the larger total surface area at the beginning, similar to cases with lower neon content. However, a reduced number of larger fragments has a longer ablation duration, resulting in a relatively similar total assimilation level when ablation ceases. In contrast to the 0.12\% neon cases, the ablation stops due to the temperature drop during the TQ, which occurs before the fragments leave the plasma on the high field side. By the end of the TQ phase, the 199 fragments case shows the highest assimilation, followed by the 1105 fragments case, with the 53 fragments case having the lowest assimilation. However, the trend is now less pronounced, and the assimilation levels are not significantly different, aligning with experimental observations that the size effect on assimilation is relatively small for pellets with higher neon fractions~\cite{stefanEPS24,paul2024arxiv}. It should be noted that, in those 10\% neon cases, a brief decline in assimilation is observed after the TQ. This is due to a transient increase in impurity diffusivity to 4 - 6 times its original value, i.e. $6.4-9.6\ \mathrm{m^2/s}$, to overcome numerical difficulties. Once this challenging period passes, the impurity diffusivity is restored to its original value. For the 199 fragments case with 10\% neon, the simulation could not proceed beyond the onset of the core collapse due to numerical issues.

        The temperature profile evolution for the 0.12\% neon cases is shown in figure~\ref{fig_tecont_size012}, and for the 10\% neon cases in figure~\ref{fig_tecont_size10}. The 10\% neon case with 199 fragments is excluded from the comparison, as this simulation did not complete the TQ. In the low neon fraction cases, an inside-out cooling pattern is observed across all configurations, owing to limited impurity assimilation at the edge. However, in the high neon fraction cases, as the number of fragments increases, the cooling pattern transitions from inside-out to outside-in. The plume generated by the smaller fragments is more elongated, ensuring a continuous presence of fragments at the edge. Additionally, smaller fragments exhibit more efficient assimilation during the initial stage. Consequently, a cold front moving alongside the pellets inwards is established more effectively with smaller fragments. This effect is clearly illustrated in figure~\ref{fig_tebelt}, where the temperature belt around 100~eV is progressively diminished as the number of small fragments increases, indicating enhanced edge cooling and a more pronounced cold front.  
     
    \subsection{Effect of penetration speed}\label{speed}
        To further study the effect of penetration speed on disruption mitigation, simulations are carried out based on the cases LF\_HV\_Ne10 and MF\_HV\_Ne10 as described in Tab.~\ref{tab1}. Generally speaking, the fragment configuration is determined by two main factors: the impact speed ($v_{\perp}$), which is the pellet velocity perpendicular to the shattering plane and influences the size of the fragments, and the penetration speed ($v_{\parallel}$), which is the pellet velocity parallel to the shattering plane and affects the velocity distribution of the fragments after shattering. To isolate the effects of these two factors, we maintain the same fragment numbers and sizes as in the cases LF\_FV\_Ne10 and MF\_FV\_Ne10 while reducing the velocities of the fragments to half of their original values (\textbf{H}alf \textbf{V}elocity). This approach retains the impact speed while altering the penetration speed. 

\begin{figure}[htb!]
\centering
\includegraphics[width=.45\textwidth]{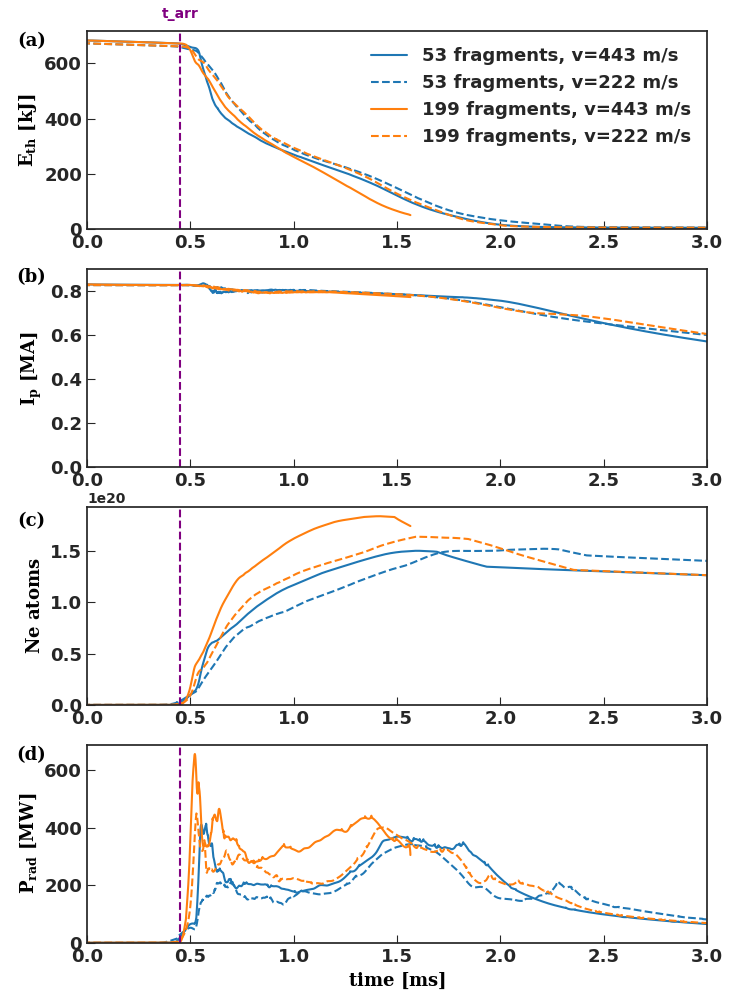}
\caption{Temporal evolution of (a) total thermal energy, (b) plasma current, (c) total assimilated neon quantity and (d) radiation power for the 53 and 199 fragment cases with different penetration speed and 10\% neon. The traces of the two slower cases are plotted with the time shifted to align the arrival time, indicated by $t_{arr}$, of the fragment with that of the faster cases.}
\label{fig_speed}
\end{figure}

\begin{figure}[htb!]
\centering
\includegraphics[width=.45\textwidth]{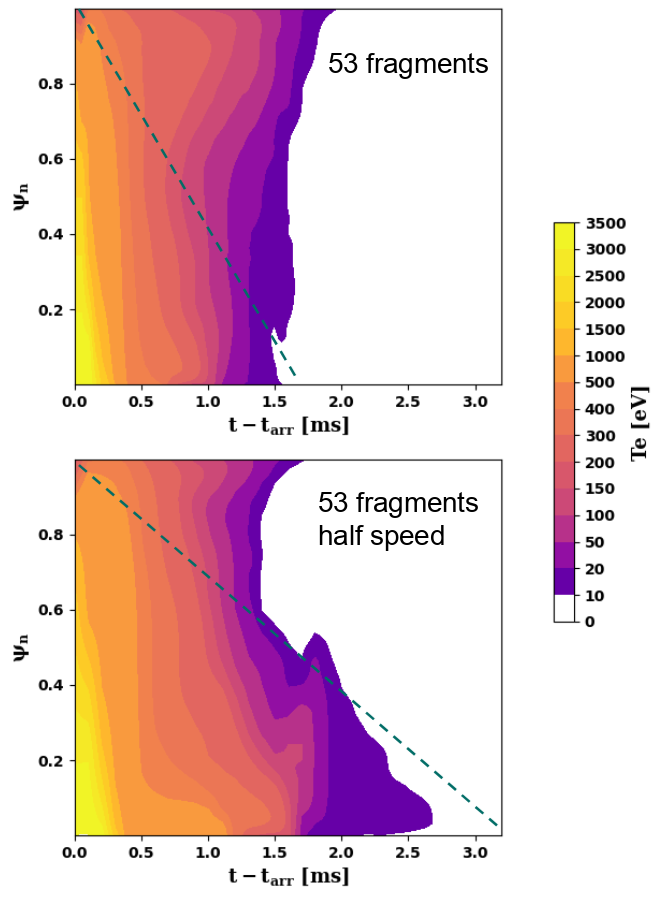}
\caption{Flux-averaged temperature profile evolution for different penetration speed with 10\% neon. The dashed lines indicate the estimated trajectory of the bulk fragments.}
\label{fig_tecont_speed}
\end{figure}

\begin{figure}[htb!]
\centering
\includegraphics[width=.45\textwidth]{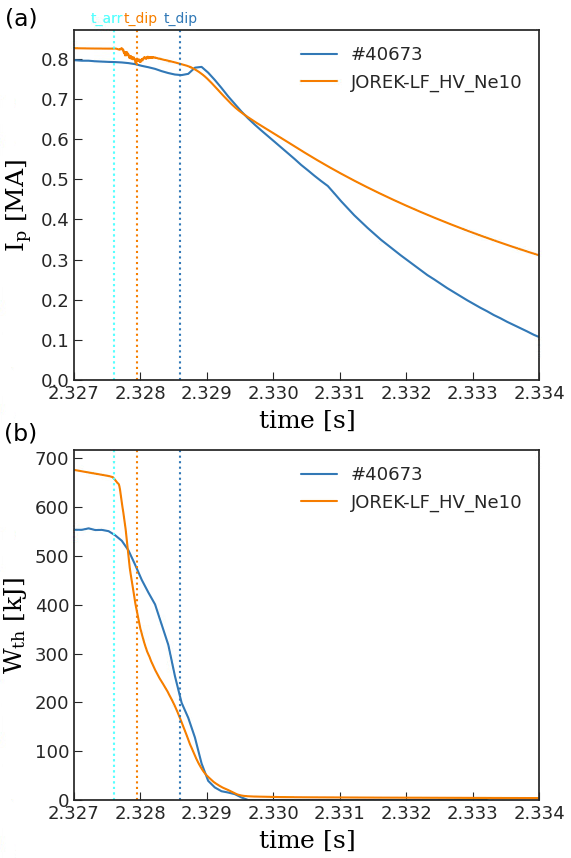}
\caption{(a) Plasma current $I_p$ and (b) thermal energy $W_{th}$ evolution for the case LF\_HV\_Ne10 in comparison with the experiment discharge \#40673. In this particular experimental discharge, the pellet speed is 221 m/s, the shatter angle is 25 degrees, and the neon fraction is 10\%. The traces of the simulation are plotted with the time shifted to align the arrival time of the fragments with experiment, which is annotated by the cyan dotted line. The orange and blue dotted lines indicate the time of $I_p$ dip for the two cases, respectively.}
\label{fig_ipspike}
\end{figure}

        The macroscopic quantities for different penetration speeds are compared in figure~\ref{fig_speed}. The traces of the two slower cases are plotted with the time shifted to align the arrival time of the fragments with that of the faster cases, for a clearer comparison. Numerical results suggest that faster penetration speeds can facilitate the material assimilation, especially during the early stage after injection, leading to shorter TQ duration. With a faster penetration speed, the fragments reach the hotter regions more quickly enhancing stronger ablation, and hence increasing the impurity radiation. The initial thermal energy drop is also more steep, comparing the slow speed cases. The transition from an inside-out to an outside-in cooling pattern can be observed with decreasing penetration speed, as shown in figure~\ref{fig_tecont_speed}. Due to the slower penetration speed, the fragments remain longer at the edge during their penetration, allowing for more efficient impurity assimilation at the edge, which contributes to the formation of the cold front.

        As the cold front forms at the edge, a more pronounced $I_p$ spike is observed in the simulation. Figure~\ref{fig_ipspike} presents the plasma current and thermal energy evolution for the case LF\_HV\_Ne10, compared with the experiment discharge \#40673. A comparable $I_p$ spike with experiment is observed in the simulation. On the other hand, the $I_p$ dip in the simulation occurs significantly earlier than that in experiment. This earlier dip is attributed to the excessively sharp thermal energy drop following the fragment arrival, as shown in the thermal energy evolution. This rapid energy loss leads to an unrealistically short pre-TQ duration, as discussed in section~\ref{neonscan}. 

    \subsection{General figures of merit}\label{trend}
        The aforementioned simulations provide valuable insights into the plasma response to various fragment parameters. In this subsection, general trends and summary plots are presented, highlighting the dependence on these parameters and how the results align with experimental observations.

\begin{figure}[htb!]
\centering
\includegraphics[width=.45\textwidth]{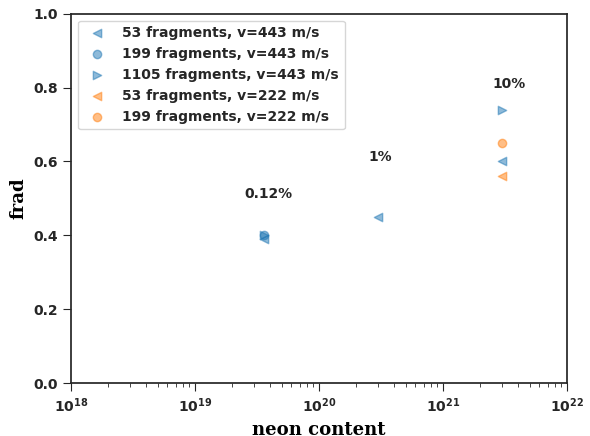}
\caption{Radiation fraction for all neon doped cases calculated by $f_{rad}=W_{rad}/(W_{th}(t=0)+W_{ohm})$ at the time point when the plasma current has dropped to $I_p=0.8I_p(t=0)$. The pure deuterium case is excluded, as neutral line radiation and background impurities are not yet considered in our current modelling.}
\label{fig_frad}
\end{figure}

\begin{figure}[htb!]
\centering
\includegraphics[width=.45\textwidth]{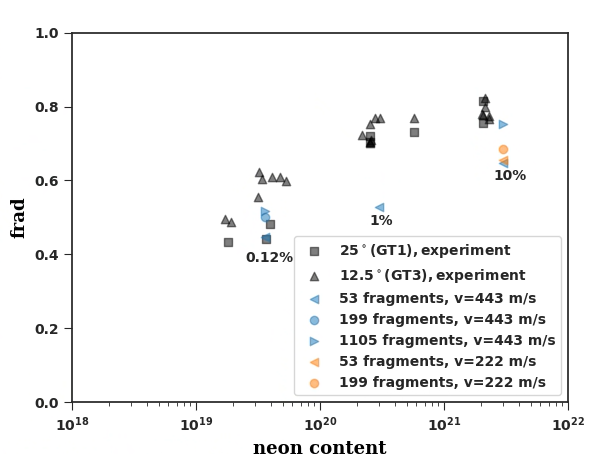}
\caption{Radiation fraction for all neon doped cases calculated by $f_{rad}=W_{rad}/(W_{th}(t=0)+W_{ohm})$ at the end of simulation, comparing with experimental results~\cite{paul2024arxiv}. The pure deuterium case is excluded, as neutral line radiation and background impurities are not yet considered in our current modelling.}
\label{fig_frad_exp2}
\end{figure}

\begin{figure*}[htb!]
\centering
\includegraphics[width=1.\textwidth]{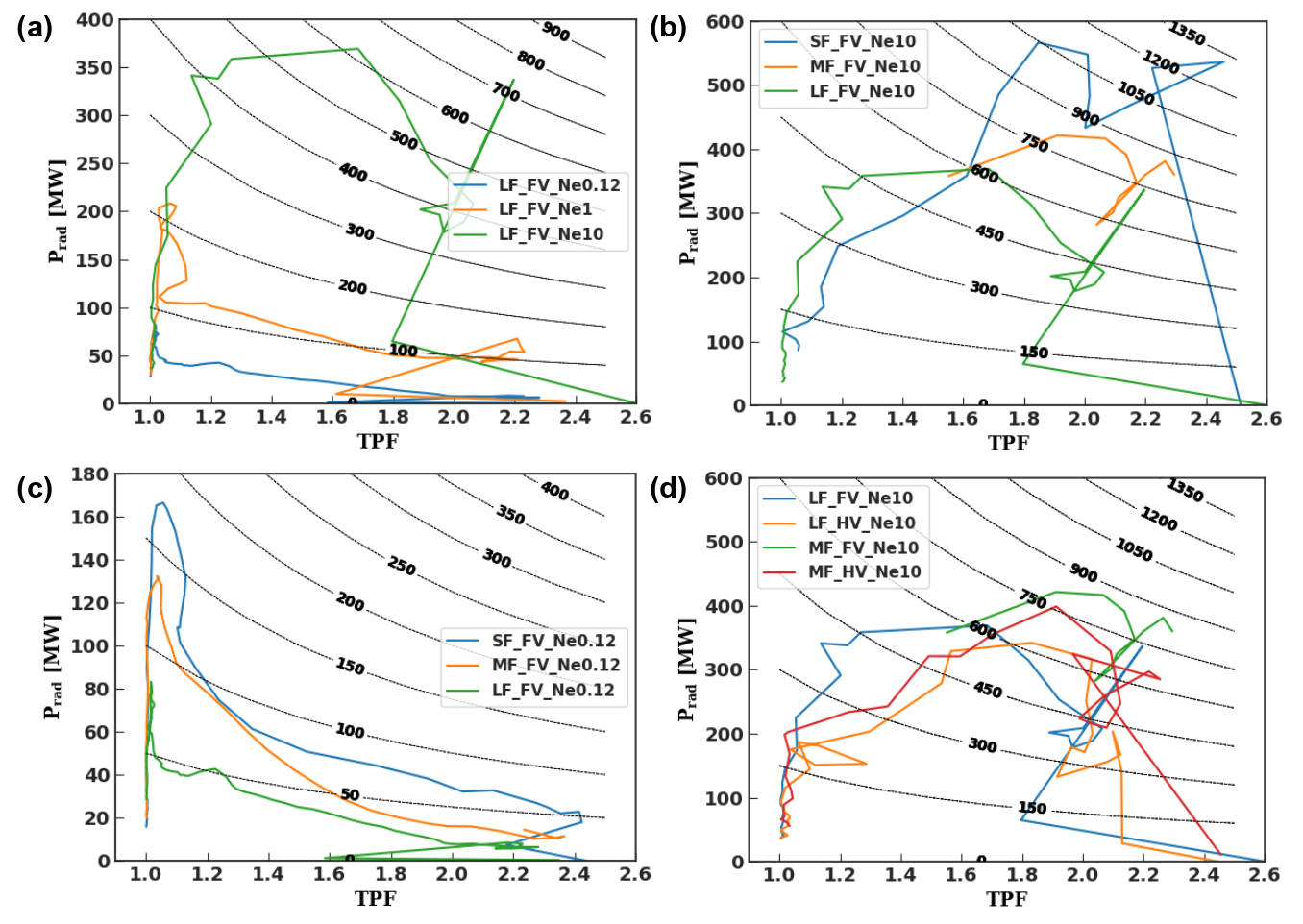}
\caption{The total radiation power vs. toroidal peaking factor for (a) different neon fraction, (b) 10\% neon with different fragment size, (c) 0.12\% neon with different fragment size and (d) different penetration speed. The TPF is calculated by $P_{rad}^{max}/\Sigma (P_{rad,i}+P_{rad,i+1})/2*\Delta_\phi(i,i+1)/360$, where $P_{rad}^{max}$ is the maximum radiation among the 5 sectors installed with foil bolometer, $(P_{rad,i}+P_{rad,i+1})/2$ is the average radiation between the neighboring sectors and $\Delta_\phi(i,i+1)$ is the toroidal angle between the two neighboring sectors. The contour plot represents the constant values of the product TPF * $P_{rad}$.}
\label{fig_rad_tpf_exp}
\end{figure*}

\begin{figure*}[htb!]
\centering
\includegraphics[width=1.\textwidth]{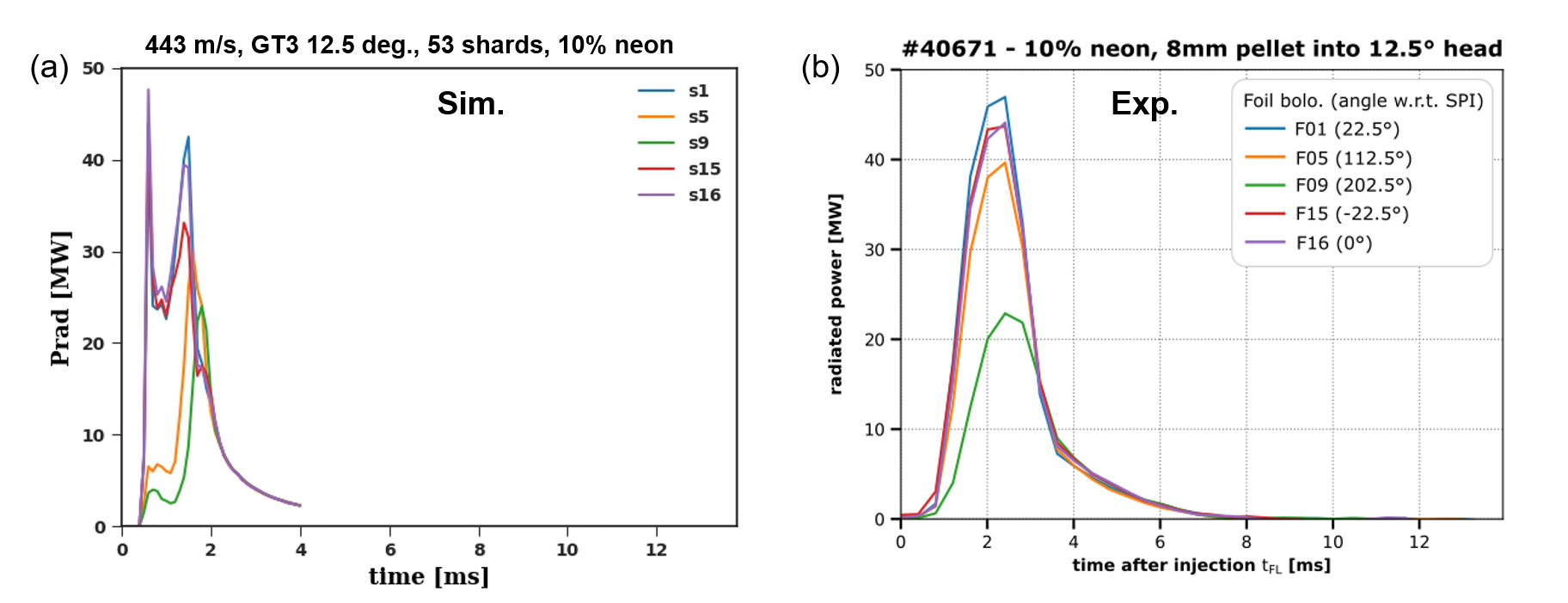}
\caption{Evolution of radiation power from different sectors for (a) the case LF\_FV\_Ne10 and its (b) counterpart in experiments. S16 is the injection sector and s1 and s15 are the neighboring sectors. S5 and S9 are sectors perpendicular and opposite to the injection sector, respectively.}
\label{fig_rad_sector}
\end{figure*}

\begin{figure}[htb!]
\centering
\includegraphics[width=.485\textwidth]{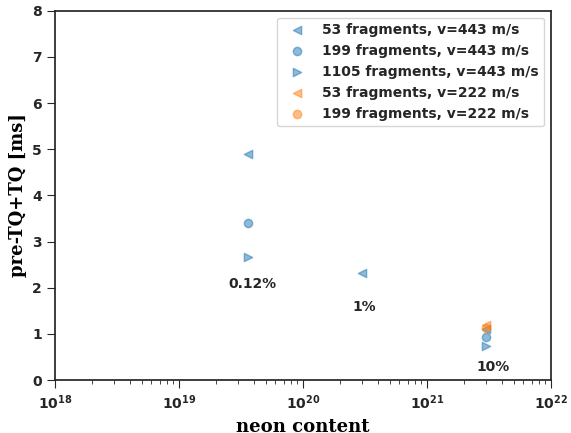}
\caption{Pre-TQ+TQ duration, period from fragment arrival to the end of TQ, for all neon doped cases. For the pure deuterium case, since a complete TQ does not occur, this duration is not available.}
\label{fig_pretq}
\end{figure}

\begin{figure}[htb!]
\centering
\includegraphics[width=.5\textwidth]{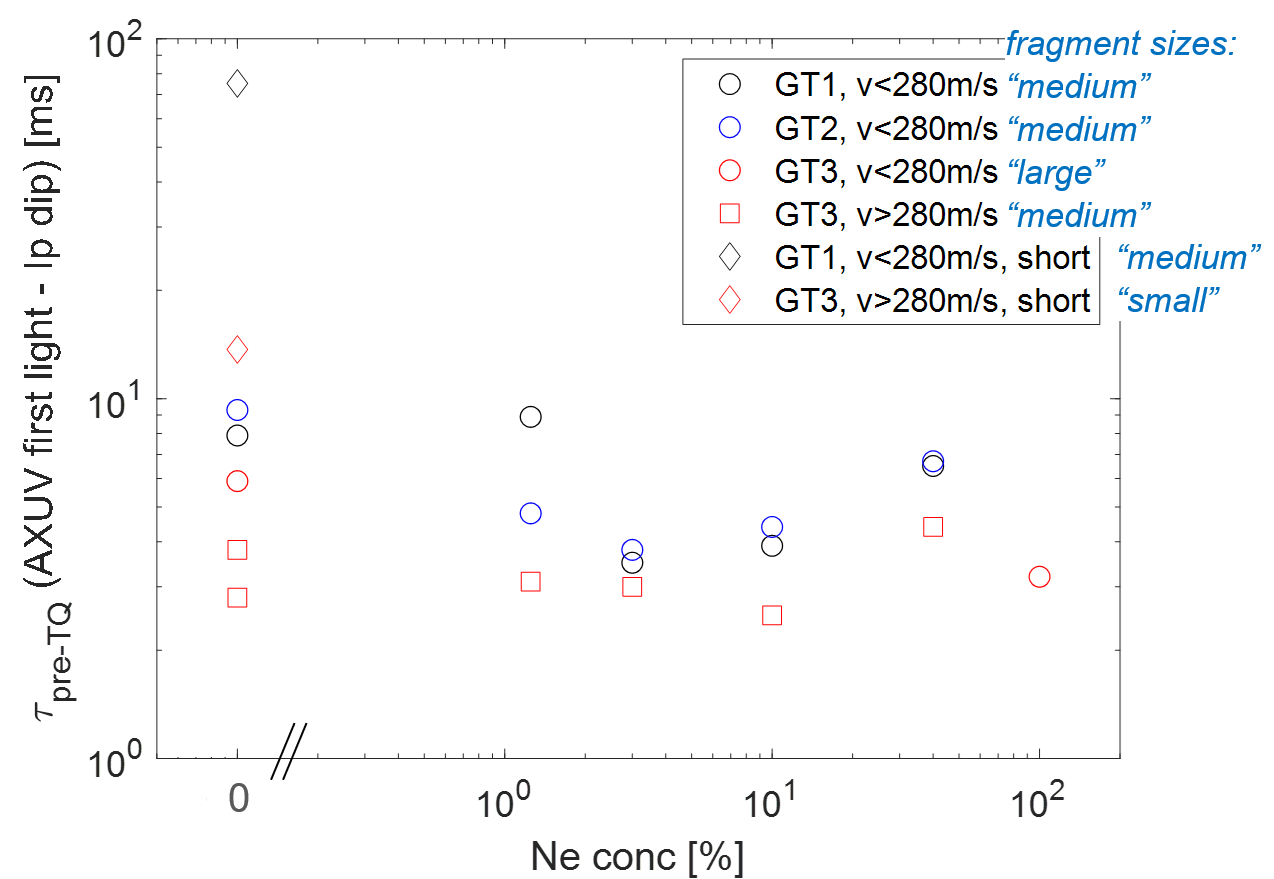}
\caption{Pre-TQ duration, period from first light $t_{FL}$ to the $I_p$ dip (proxy for the TQ onset), in AUG experiments.}
\label{fig_pretq_exp}
\end{figure}

\begin{figure}[htb!]
\centering
\includegraphics[width=.5\textwidth]{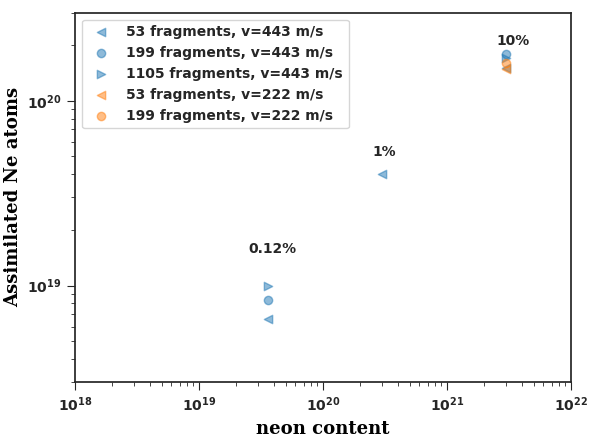}
\caption{Total assimilated neon atoms for all neon doped cases with different neon fractions, fragment sizes and speeds.}
\label{fig_assne}
\end{figure}

\begin{figure}[htb!]
\centering
\includegraphics[width=.5\textwidth]{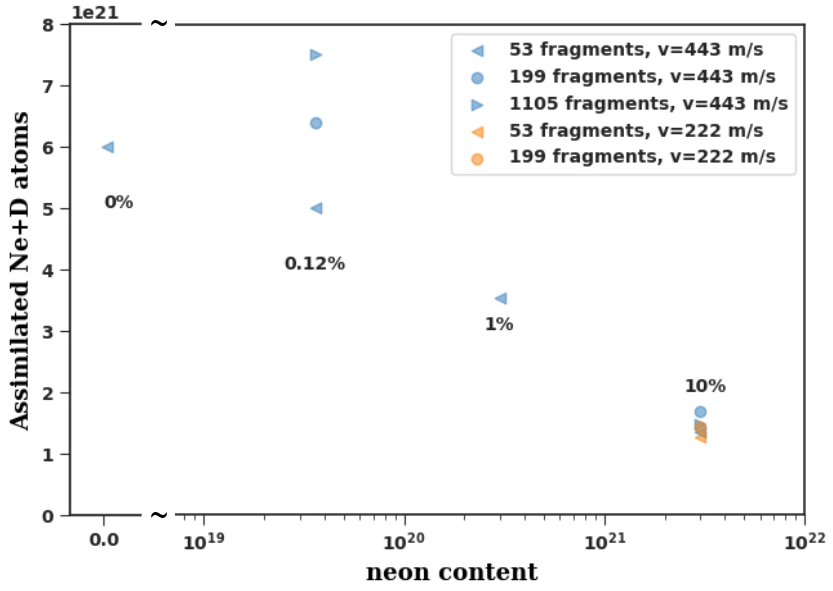}
\caption{Total assimilated neon and deuterium atoms for all the cases with different neon fractions, fragment sizes and speeds.}
\label{fig_asstot}
\end{figure}
        
        Figure~\ref{fig_frad} illustrates the radiation fraction similar to the experimental definition following Ref.~\cite{paul2024arxiv}  $f_{rad}=W_{rad}/(W_{th}(t=0)+W_{ohm})$ at the time point when the plasma current $I_p$ has dropped to 80\% of its original value for a consistent comparison, since not all the cases reach at the same CQ level by the end of simulation. Here, $W_{rad}$ is the total radiated energy, $W_{th}(t=0)$ is the thermal energy at the beginning, and $W_{ohm}$ is the total ohmic heating energy in the considered time interval. It should be mentioned that, in our current modeling, as described in section~\ref{sec2}, the deuterium neutral fluid and neutral line radiation are not considered. Additionally, no background impurities are included in the 3D simulations. Consequently, for the pure deuterium case, the radiation fraction is 0\%. As for neon-doped pellets, the radiated energy accounts for approximately 40\% of the lost energy for the 0.12\% neon case, and 45\% for the 1\% neon case. Furthermore, the size of the fragments does not have any significant impact on the radiation fraction in the 0.12\% neon cases, owing to the low radiative cooling effect during the first stage of thermal energy loss. For 10\% neon cases, the radiation fraction ranges from 56\% - 74\%, depending on different injection scenarios. Faster and smaller fragments exhibit better performance in terms of radiation fraction, as they generate higher radiation during the first stage of thermal energy loss. 

        To compare with experiments, the radiation fraction, calculated using the above definition at the end of each simulation, is plotted alongside experimental data in figure~\ref{fig_frad_exp2}. The experimental data is derived until the end of the CQ~\cite{paul2024arxiv}. The overall radiation fraction in the simulations is lower than that observed in experiments. This discrepancy is primarily due to excessive thermal energy loss during the first stage of the cooling process, which is characterized by a low radiation level. The key to improve the radiation fraction lies in increasing radiative cooling during the first stage. Notably, the case with 10\% neon and 1105 fragments shows a comparable radiation fraction to experimental results, owing to more effective radiative coverage in the initial cooling phase. 

        The total radiation power versus the toroidal peaking factor (TPF) is plotted in figure~\ref{fig_rad_tpf_exp}, where the TPF is obtained using the same formula in experiments~\cite{PaulAPS24}. The key message here is that there are two distinct radiation peaks, which are much more prominent in the large neon fraction cases. The first peak occurs at the initial injection and the second at the end of the TQ. The first peak, characterized by a relatively large TPF, is mainly caused by the outward heat flux that result in sharp ablation of the fragments during the first stage of thermal energy drop, which is overestimated in simulations and not observed in experiments. The smaller fragment case exhibits a stronger TPF and higher radiation at this stage. The TPF at the TQ is generally very small for all cases, as the ablated material has spread spatially at the latter phase, which is also observed experimentally. It is crucial to prevent high toroidal peaking during the initial injection phase in experiments, as simulations suggest that this phase is more likely to produce a high TPF, despite the fact that our current simulations tend to overestimate the TPF at this stage. One possible explanation for the more peaked early-phase radiation peaking observed in our simulations is opacity. During the early phase, the local density around the fragments could become very high so that the plasma is not transparent for line radiation, reducing the local radiation power density around the fragments in reality~\cite{Aleynikov_2024,Runov_Aleynikov_Arnold_Breizman_Helander_2021}.

        In addition, the distribution of radiation across different toroidal sectors in the experiment is more uniform (except S9) compared to what was observed in simulation, as illustrated in figure~\ref{fig_rad_sector}. In the simulation, the difference in radiation between sectors near the injection plane and those far from it is more significant, comparing with experiments~\cite{PaulAPS24}. This may be due to the fact that the equilibrium toroidal flow is not considered in the simulation.

        Figure~\ref{fig_pretq} presents the pre-TQ+TQ duration for all neon-doped cases. In the case of pure deuterium injection, no complete TQ is observed due to the lack of radiative cooling, resulting in an effectively infinite TQ duration. This behavior aligns with non-disruptive shots or those with extended pre-TQ durations seen in experiments following pure deuterium injection~\cite{paul2024arxiv}. For neon-doped injections, even a slight neon fraction of 0.12\% reduces the pre-TQ+TQ time to 3 – 5~ms. As the neon fraction increases to 10\%, this duration further decreases to approximately 1~ms. Additionally, for cases with low neon fractions, the pre-TQ+TQ time shows a noticeable dependence on fragment size, which can be attributed to the significant differences in material assimilation in our simulations.

        Figure~\ref{fig_pretq_exp} illustrates the correlation between pre-TQ duration and neon fraction observed in AUG experiments. In the pure deuterium cases, the pre-TQ durations vary significantly, ranging from a few milliseconds to tens of milliseconds. However, once the neon fraction exceeds 1\%, the pre-TQ durations tend to converge, with most cases falling between 3 to 5~ms. This trend also suggests that JOREK simulations presented in this article underestimate the pre-TQ duration, especially with a low neon fraction, because of the exaggerated initial thermal energy drop. 

        In terms of the assimilation, figure~\ref{fig_assne} exhibits the assimilated neon atoms for all the neon doped cases. As expected, neon assimilation in absolute numbers generally increases with higher neon fractions. For the 0.12\% neon cases, there is a clear dependence on the fragment size, with smaller fragments showing higher assimilation, which deviates from the experimental results. This is because our simulations in this work fail in accurately simulating the pre-TQ duration, as we discussed in section~\ref{size}. On the other hand, for cases with higher neon fractions, such as the 10\% neon case, the dependence on fragment size is less pronounced due to the extended ablation duration for larger fragment case. With respect to penetration speed, faster fragments result in higher assimilation during the early stage after injection, as shown in figure~\ref{fig_speed}; however, the difference is very small in the end.

        Regarding the total assimilation, figure~\ref{fig_asstot} displays the total assimilated atoms (neon + deuterium) for all cases. The trend reverses when considering the total assimilated atoms, with total assimilation increasing as the neon fraction decreases. This is due to the lower assimilation ratio observed in high neon fraction cases, primarily because of the much shorter TQ duration. As these pellets contain comparable amounts of deuterium, and deuterium atoms far outnumber neon atoms within the pellet, a higher ablation ratio leads to enhanced deuterium assimilation. The dependence on fragment size and speed follows a similar pattern to that of neon assimilation.
        
\section{Summary and discussion}\label{sec4}
    In this study, 3D non-linear SPI simulations for an AUG H-mode scenario are carried out using realistic parameters with the 3D non-linear JOREK code, aimed at validating optimal fragment configurations for the ITER DMS. The key findings can be summarized as follows:

    Two distinct cooling phases are identified in the simulations. The first is dominated by convective and conductive transport caused by magnetic field stochastization at the plasma edge, virtually independent of injection parameters. The second phase, more prominent in neon-doped cases, is driven by radiative cooling, which is critical for the TQ duration. As for pure deuterium injection without background impurity, the second phase of cooling may not occur at all due to limited radiation, corresponding to non-disruptive shots observed in experiments. Results show that high neon fraction (10\%) leads to faster TQ through radiative processes, while lower fraction (0.12\%) result in more gradual cooling. High neon fraction injections (10\%) cause stronger MHD response during the second stage cooling, leading to faster core collapse and more intense cooling. Low neon fractions slow down this process, leading to longer TQ durations and a more gradual loss of thermal energy.

    The total assimilation becomes less efficient with increasing neon content. Higher neon fraction can lead to shorter ablation duration, resulting in a lower ablation ratio. For the low neon fraction (0.12\%) cases, the assimilation exhibits a strong dependence on fragment size. Smaller fragments increase the surface area available for ablation, leading to higher assimilation. This trend diverges from experimental observations, where larger fragments typically lead to higher material assimilation. This discrepancy may stem from inaccuracies in simulating the pre-TQ phase. It is possible that a larger number of smaller fragments could enhance plasmoid drift due to quicker ablation and increased local pressure peaking. The plasmoid drift dynamics in JOREK is systematically investigated in Ref.\cite{plasmoid_kong}, by considering an instantaneous massive gas injection source, and while the results show qualitative agreement with theoretical expectations, quantitative modeling remains very challenging due to limited toroidal resolution. Coupling analytical model or dedicated code on pellet physics could be a promising path forward for incorporating this physics in a more self-consistent way\cite{Samulyak_2021,Akinobu_2022}. Additionally, the current model does not incorporate the rocket effect, which might be particularly important for small fragments, as their smaller mass makes them more susceptible to this effect. For the cases with high neon fraction (10\%), the dependence of assimilation on fragment size is less pronounced. The smaller fragments lead to shorter ablation duration, even though the ablation rate is higher. This part of results aligns with experimental observations, likely due to the suppression of plasmoid drift at high neon fraction injections in experiments.

    In cases involving large fragments (53 fragments), an inside-out cooling pattern emerges due to less efficient edge cooling. The results suggest that smaller and slower fragments promote edge cooling, which aids in forming a cold front and shifting the cooling pattern from inside-out to outside-in. The plume generated by smaller fragments is more elongated, ensuring a continuous presence of fragments at the edge. Additionally, smaller fragments demonstrate more efficient assimilation during the initial stages, while slower fragments remain longer at the edge during penetration, further enhancing edge cooling. As the cold front forms at the edge, an $I_p$ spike comparable to the experimental one is observed in the simulation.

    The overall radiation fractions observed in the simulations are slightly lower than those seen in experiments, likely due to overestimated thermal energy loss in the first stage of cooling. This discrepancy is primarily due to excessive thermal energy loss during the first stage of cooling, which is not sufficiently radiated. This is likely due to the overestimation of the parallel thermal conduction in our simulations. There are two major ways in which an overestimated thermal conduction can artificially enhance MHD activity in the simulation. First, excessive thermal conduction promotes stronger cooling along magnetic field lines, which leads to a rapid increase in plasma resistivity. This enhanced resistivity, in turn, facilitates the growth of resistive MHD instabilities, thereby exaggerating the MHD response. Secondly, in regions where the magnetic topology becomes stochastic, the overestimated thermal conduction results in unrealistically high thermal energy losses, especially in the initial cooling phase. This excessive energy depletion leads to a shorter and more abrupt pre-TQ phase than what is observed in experiments. As for the radiation asymmetry, two radiation peaks are observed in large neon fraction cases, one at initial injection and another at the end of the TQ. The first peak is caused by the outward heat flux during the initial thermal energy drop. The smaller fragment case shows stronger peaking factors and higher radiation during this stage. The radiation peak at TQ shows a small peaking factor due to a more relaxed plasma state. The radiation distribution in experiments is more uniform across toroidal sectors than in simulations, likely because our current simulations do not include the equilibrium toroidal flow and the opacity effect. The current work aims to validate the SPI model in JOREK and compare the general trends with experimental results. More detailed comparisons with experiments will be carried out in the future work.  
    
\section*{Acknowledgements}
    We sincerely thank Sang-Jun Lee, Mengdi Kong, Daniele Bonfiglio, Andres Cathey, Nina Schwarz and Haowei Zhang for fruitful discussions. We are also very grateful for the CPU time generously provided by the HAWK supercomputer operated by the High Performance Computing Center (HLRS) in Stuttgart, Germany. Part of this work has been carried out within the framework of the EUROfusion Consortium, funded by the European Union via the Euratom Research and Training Programme (Grant Agreement No 101052200 — EUROfusion). Within EUROfusion, in particular the Theory and Simulation Verification and Validation (TSVV) projects on MHD Transients (PI M. Hoelzl) and the European HPC infrastructure (Marconi-Fusion) provided essential support. Views and opinions expressed are however those of the author(s) only and do not necessarily reflect those of the European Union or the European Commission. Neither the European Union nor the European Commission can be held responsible for them. Part of this work has been carried out in collaboration with the ITER Organization. The views and opinions expressed herein do not necessarily reflect those of the ITER Organization. This work receives funding from the ITER Organization under contract IO/20/IA/43-2200. The ASDEX-Upgrade SPI project has been implemented as part of the ITER DMS Task Force programme. The SPI system and related diagnostics have received funding from the ITER Organization under contracts IO/20/CT/43-2084, IO/20/CT/43-2115, IO/20/CT/43-2116.
\section*{References}
\bibliography{paper}
\end{document}